\documentclass[11pt]{article}
\usepackage{amsmath,bm}
\usepackage{amssymb}
\usepackage{geometry}
\usepackage{graphicx}
\usepackage{hyperref}
\usepackage{subfig}
\usepackage{mciteplus}

\def\gtwid{\mathrel{\raise.3ex\hbox{$>$\kern-.75em\lower1ex\hbox{$\sim
$}}}}
\def\vio{\mathrel{\hbox{$E$\kern-.60em\hbox{$/
$}}}}
\def\lsim{\mathrel{\raise.3ex\hbox{$<$\kern-.75em\lower1ex\hbox{$\sim$}}}}
\def\gsim{\mathrel{\raise.3ex\hbox{$>$\kern-.75em\lower1ex\hbox{$\sim$}}}}

\newcommand{\gev}{\text{GeV}}

\newcommand{\m}{\text{m}}

\newcommand{\mchi}{m_{\chi}}

\newcommand{\be}{\begin{equation}}
\newcommand{\ee}{\end{equation}}
\newcommand{\bea}{\begin{eqnarray}}
\newcommand{\eea}{\end{eqnarray}}

\newcommand*{\signmu}{\ensuremath{\text{sgn}\,\mu}}

\begin{document}

\title
{Bayesian implications of current LHC supersymmetry and dark matter detection
  searches for the Constrained MSSM} 
\author{Leszek Roszkowski\footnote{On leave of absence from the University of
Sheffield.},~Enrico Maria Sessolo and Yue-Lin Sming Tsai\\[2ex]
\small\it National Centre for Nuclear Research, Ho$\dot{z}$a 69, 00-681 Warsaw, Poland\\
}

\date{}

\maketitle

\begin{abstract}
  We investigate the impact of recent limits from LHC searches for
  supersymmetry and from direct and indirect searches for dark matter
  on global Bayesian inferences of the parameter space of the
  Constrained Minimal Supersymmetric Standard Model (CMSSM). In particular we apply recent exclusion limits
  from the CMS $\alpha_T$ analysis of 1.1~fb$^{-1}$ of
  integrated luminosity, the current direct detection dark matter limit from
  XENON100, as well as recent experimental constraints on $\gamma$-ray
  fluxes from dwarf spheroidal satellite galaxies of the Milky Way
  from the FermiLAT telescope, in addition to updating values for other
  non-LHC experimental constraints. We extend the range of scanned
  parameters to include a significant fraction of the focus
  point/hyperbolic branch region. While we confirm earlier conclusions
  that at present LHC limits provide the strongest constraints on the
  model's parameters, we also find that when the uncertainties are not
  treated in an excessively conservative way, the new bounds from dwarf spheroidals
 have the power to significantly constrain the focus
  point/hyperbolic branch region. Their effect is then comparable, if
  not stronger, to that from XENON100. We further analyze the effects
  of one-year projected sensitivities on a neutrino
  flux from the Sun in the 86-string IceCube+DeepCore configuration
  at the South Pole. We show that data on neutrinos from the Sun,
  expected for the next few months at IceCube and DeepCore, have the
  potential to further constrain the same region of parameter space independently of the LHC
  and can yield additional investigating power for the model.
\end{abstract}

\newpage

\section{Introduction}\label{intro:sec}

The Large Hadron Collider (LHC) at CERN started its run at center of
mass energy $\sqrt{s}=7$~TeV more than a year ago, and in 2011
published several results based on an integrated luminosity of
1.1~fb$^{-1}$~\cite{Chatrchyan:2011zy,arXiv:1109.6572}.  Several
searches were performed, aimed at the discovery of signatures
that might be an indication of new physics beyond the Standard Model
(SM). One of the scenarios that is most appealing and well motivated
from a theoretical point of view is softly broken
supersymmetry~(SUSY) (see, e.g., Ref.~\cite{Baer:2006rs}). In the past
25 years or so there has been vast proliferation of models
for SUSY-breaking. Among models that are mediated by gravitational
interactions, the Constrained Minimal Supersymmetric Standard
Model~(CMSSM)~\cite{hep-ph/9312272} is the most popular because of
reasonable unification assumptions and a reduced number of
parameters. In the CMSSM, three parameters are defined at the scale of
grand unification: the universal scalar mass $m_0$, the universal
gaugino mass $m_{1/2}$, and the universal trilinear coupling
$A_{0}$. Additionally, $\tan\beta$ (the ratio
of the expectation values of the two Higgs doublets) is defined at
the electroweak scale, while the sign of the Higgs/higgsino parameter
$\mu$ remains undetermined.

Recently, there have been many studies, both experimental and
phenomenological, of the impact of bounds from new physics searches
at the LHC on the parameter space~(PS) of the CMSSM~\cite{Buchmueller:2011aa,*Buchmueller:2011ki,*Allanach:2011ut,*Allanach:2011wi,*Farina:2011bh,Buchmueller:2011sw,
Akula:2011dd,Bertone:2011nj,*Strege:2011pk}.
In particular, a recent $\alpha_T$ analysis~\cite{Chatrchyan:2011zy}
excluded for most MSSM models observation of the lightest squarks with
mass $m_{\tilde{q}}\lesssim 700$~GeV, and gluinos with
$m_{\tilde{g}}\lesssim 580$~GeV at the 95\% confidence level
(C.L.). Under sensible assumptions, these limits translate into an
exclusion bound in the ($m_0$, $m_{1/2}$) plane at fixed $A_0=0$,
$\tan\beta=10$. The bound is shown in Fig.~8 of
Ref.~\cite{Chatrchyan:2011zy}.

Recently, an updated Bayesian analysis of the CMSSM, which included the
$\alpha_T$ bound from CMS, was performed by some of the authors of this
paper~\cite{Fowlie:2011mb}. The analyzed region covered the range
$m_0\leq2$~TeV, $m_{1/2}\leq1$~TeV, and also included the improved
limits on the spin-independent~(SI) neutralino-proton cross section
$\sigma_p^{\textrm{SI}}$ from direct detection~(DD) of dark
matter~(DM) with 100.9 days of live data at
XENON100~\cite{arXiv:1107.2155,Aprile:2011hx,arXiv:1104.2549}. One of
the main findings of~\cite{Fowlie:2011mb} was that LHC
limits currently have the most constraining effect on the PS of the CMSSM, while
a further impact of the DD limit from XENON100 is limited after
theoretical and astrophysical uncertainties are taken into account.

The study~\cite{Fowlie:2011mb} also showed that a combination of the
new LHC limits with the previously considered usual non-LHC
constraints, particularly from flavor physics [$b\rightarrow s\gamma$,
$(g-2)_{\mu}$, $B_s\rightarrow \mu^+\mu^-$, etc.], dark matter abundance $\Omega_{\chi}h^2$, and LEP, results in two
broad regions of high posterior probability. One is located at large
$m_{1/2}$ and small $m_0$ and agrees reasonably well with the
findings of studies using a frequentist approach~\cite{Buchmueller:2011sw}. In
addition, a sizable region remains in the focus
point/hyperbolic branch (FP/HB) region~\cite{Baer:2006rs,Feng:1999zg,*Feng:2000gh,*Chan:1997bi}, the part of PS
at large $m_{0}$, and $m_{1/2}<m_0$.

In this paper we look deeper into the FP/HB region and perform a
Bayesian analysis of the CMSSM with its soft mass parameters extended
to $m_0\le 4$~TeV and $m_{1/2}\le 2$~TeV, which is for each parameter twice as large as in~\cite{Fowlie:2011mb}. While generally inaccessible to direct
LHC searches for SUSY, the FP/HB region can be probed by results from
DD of DM, and further constraints can be derived from recent results
on indirect detection (ID).  In that respect, in addition to updated
constraints used in~\cite{Fowlie:2011mb}, we apply the recent upper
bounds on the annihilation cross section of neutralinos obtained by
the FermiLAT Collaboration~\cite{arXiv:1108.3546} from $\gamma$-ray
fluxes from dwarf spheroidal satellite galaxies of the Milky Way
(dSphs). Additionally, we discuss the impact of the one-year
95\%~C.L. sensitivities for observation of high energy neutrinos from
the Sun at IceCube and DeepCore~\cite{arXiv:1111.5188}.

A full list of constraints applied here and also
in~\cite{Fowlie:2011mb} will be given below. Recently, the impact on
the CMSSM parameters of
LHC limits and DD (but not ID)  limits on DM 
was also investigated in
Refs.~\cite{Buchmueller:2011sw,Akula:2011dd,Bertone:2011nj,*Strege:2011pk}.
Reference~\cite{Buchmueller:2011sw} adopted frequentist
statistics and~\cite{Akula:2011dd} performed a fixed-grid scan
in contrast to our Bayesian one, while
Refs.~\cite{Bertone:2011nj,*Strege:2011pk} included a different set of
constraints. In contrast, the main goal of this paper is to analyze the impact
of ID searches.


In Sec.~\ref{methodology:sec} we summarize the main features of Bayesian inference and
describe our scanning methodology. We explain our implementation of
the LHC and XENON100 constraints in Sec.~\ref{lhcdd:sec}. In Sec.~\ref{id:sec} we describe our
implementation of the bounds from dSphs into the CMSSM PS, and detail
our calculation of the IceCube/DeepCore sensitivities. We discuss our
results in Sec.~\ref{results:sec}, and finally give our summary and conclusions in
Sec.~\ref{summary:sec}.

\section{Methodology}\label{methodology:sec}

In Bayesian statistics, for a theory described by some parameters $m$, experimental observables
$\xi(m)$ can be compared with data $d$ and a posterior probability density
function (pdf) $p(m|d)$ can be calculated through Bayes' Theorem
\begin{equation}
p(m|d)=\frac{p(d|\xi(m))\pi(m)}{p(d)}\, ,
\label{Bayesth}
\end{equation}
where the likelihood $p(d|\xi(m))$ gives the probability density for obtaining $d$ from a
measurement of $\xi$, the prior $\pi(m)$ parametrizes
assumptions about the theory prior to performing the measurement, and the
evidence $p(d)$ represents the assumptions on the
data. As long as one considers only one model (as
we do in this paper) it is a constant in theory parameters and thus a
normalization factor.  The Bayesian approach yields a simple and
natural procedure for calculating the posterior pdf of any
limited subset of $r$ variables in PS, $\psi_{i=1,..,r}\subset m$.
One just needs to marginalize, or integrate, over the remaining
parameters
\begin{equation}
p(\psi_{i=1,..,r}|d)=\int p(m|d)d^{n-r}m\,,
\label{marginalization}
\end{equation}
where $n$ denotes the dimension of the full PS. To describe our
methodology for the Bayesian scan, we use the same notation
as~\cite{Fowlie:2011mb}.

We scan over the CMSSM parameters $(m_0,m_{1/2},A_0,\tan\beta$), which
are incorporated in the likelihood function according to the procedure
detailed in~\cite{deAustri:2006pe}. The Bayesian approach also allows a
consistent treatment of residual uncertainties in the SM. In the
CMSSM, the relevant ``nuisance" parameters are~\cite{deAustri:2006pe} the pole mass of the
top quark $m_t^{\rm pole}$, the mass of the bottom quark at its
nominal scale in the $\overline{MS}$ scheme,
$m_b(m_b)^{\overline{MS}}$, and the strong and inverse electromagnetic
coupling constants at the $Z$ pole in the $\overline{MS}$ scheme,
$\alpha_s(M_Z)^{\overline{MS}}$ and
$1/\alpha_{em}(M_Z)^{\overline{MS}}$, respectively.

We perform our scan in the following ranges:
\begin{eqnarray}
100\textrm{ GeV}&\leq m_0\leq&4000\textrm{ GeV}\nonumber\\
100\textrm{ GeV}&\leq m_{1/2}\leq&2000\textrm{ GeV}\nonumber\\
-2000\textrm{ GeV}&\leq A_0\leq&2000\textrm{ GeV}\nonumber\\
3&\leq \tan\beta\leq&62\,.
\end{eqnarray}
\signmu\ is fixed at +1 to accommodate the discrepancy of 
$(g-2)_{\mu}$ with SM predictions~\cite{arXiv:0809.3792}. The nuisance
parameters are scanned in the ranges
 \begin{eqnarray}
163.7\textrm{ GeV}&\leq m_t^{\rm pole}\leq&178.8\textrm{ GeV}\nonumber\\
3.92\textrm{ GeV}&\leq m_b(m_b)^{\overline{MS}}\leq&4.48\textrm{ GeV}\nonumber\\
0.1096&\leq \alpha_s(M_Z)^{\overline{MS}}\leq&0.1256\nonumber\\
127.846&\leq 1/\alpha_{em}(M_Z)^{\overline{MS}}\leq&127.99\,.
\end{eqnarray}
We adopt logarithmic priors in $m_0$ and $m_{1/2}$ throughout our
analysis, and we refer the reader to the relevant
literature~\cite{deAustri:2006pe,arXiv:0809.3792} for a discussion on
the incidence of linear priors on SUSY PS scans. The observables we
use to constrain the PS, and their experimental and theoretical errors,
are reported in Table~\ref{tab:exp_constraints} and are the same as
given in Ref.~\cite{Fowlie:2011mb}. We perform our global scan with
the package SuperBayeS~\cite{SuperBayeS}, which we have modified to
our needs. As stated above, the non-LHC constraints have been updated
with respect to previous papers. To these we add the LHC, DD, and ID
constraints, which we describe in some detail below.

\begin{table}
\begin{center}
\small\addtolength{\tabcolsep}{-5pt}
\begin{tabular}{|cccccc|}
\hline \hline
Observable & Mean & Exp.~Error & Theor.~Error & Likelihood Distribution & Reference\\
\hline\hline
\multicolumn{6}{|l|}{Non-LHC:} \\
\hline
$\Omega_{\chi}h^2$          & $0.1120$  & $0.0056$      & $10\%$        & Gaussian &  \cite{Komatsu:2010fb}\\
$\sin^2\theta_{\textrm{eff}}$           & $0.231160$    & $0.00013$     & $15.0 \times 10^{-5}$ & Gaussian &  \cite{Nakamura:2010zzi}\\
$M_W$                       & $80.399$      & $0.023$       & $0.015$               & Gaussian &  \cite{Nakamura:2010zzi}\\
$\delta(g-2)^{SUSY}_{\mu}\times 10^{10}$    & $30.5 $   & $8.6$     & $1$           & Gaussian &  \cite{Nakamura:2010zzi,Miller:2007kk} \\
$\mathcal{BR}(\bar{B}\rightarrow X_s\gamma)\times 10^{4}$       & $3.6$     & $0.23$    & $0.21$        & Gaussian &  \cite{Nakamura:2010zzi}\\
$\mathcal{BR}(B_u\rightarrow\tau\nu) \times 10^{4}$          & $1.66$   & $0.66$    & $0.38$        & Gaussian &  \cite{Asner:2010qj}\\
$\Delta M_{B_{s}}$                         & $17.77$    & $0.12$    & $2.4$         & Gaussian &  \cite{Nakamura:2010zzi}\\
$\mathcal{BR}(B_s\rightarrow\mu^+\mu^-)$            & $< 1.5 \times 10^{-8}$    & $0$ & $14\%$      & Upper limit --- Error fn &  \cite{Serrano:2011}\\
\hline
\multicolumn{6}{|l|}{LEP --- $95$\% Limits} \\
\hline
$m_h$               & $>114.4$           & $0$& $3$   & Lower limit --- Error fn & \cite{Barate:2003sz} \\
$\zeta_h^2$         & $<f\left(m_h\right)$   & $0$& $0$   & Upper limit --- Step fn  & \cite{Barate:2003sz} \\
$m_{\chi}$          & $>50$              & $0$& $5$\% & Lower limit --- Error fn & \cite{Heister:2003zk} (\cite{Heister:2001nk,Achard:2003ge}) \\
$m_{\chi^\pm_1}$    & $>103.5$  ($92.4$)     & $0$& $5$\% & Lower limit --- Error fn & \cite{lepsusy} (\cite{Heister:2001nk,Achard:2003ge}) \\
$m_{\tilde{e}_R}$   & $>100$    ($73$)       & $0$& $5$\% & Lower limit --- Error fn & \cite{lepsusy} (\cite{Heister:2001nk,Achard:2003ge})  \\
$m_{\tilde{\mu}_R}$     & $>95$     ($73$)       & $0$& $5$\% & Lower limit --- Error fn & \cite{lepsusy} (\cite{Heister:2001nk,Achard:2003ge})  \\
$m_{\tilde{\tau}_1}$    & $>87$     ($73$)       & $0$& $5$\% & Lower limit --- Error fn & \cite{lepsusy} (\cite{Heister:2001nk,Achard:2003ge}) \\
$m_{\tilde{\nu}}$   & $>94$     ($43$)       & $0$& $5$\% & Lower limit --- Error fn & \cite{Abdallah:2003xe} (\cite{Nakamura:2010zzi}) \\
\hline\hline
\multicolumn{6}{|l|}{LHC CMS $\alpha_T$ 1.1/fb analysis} \\
\hline
$\alpha_T$      & See text  & See text      & $0$   & Poisson  &\cite{Chatrchyan:2011zy}\\
\hline\hline
\multicolumn{6}{|l|}{XENON100} \\
\hline
$\sigma_p^{\textrm{SI}}\left(m_{\chi} \right)$    & $< f\left(m_{\chi} \right)$ --- see text     & $0$   & $1000\%$  & Upper limit --- Error fn  &\cite{arXiv:1104.2549}\\
\hline\hline
\multicolumn{6}{|l|}{Nuisance} \\
\hline
$1/\alpha_{em}(M_Z)^{\overline{MS}}$            & $127.916$ & $0.015$ & $0$ & Gaussian &  \cite{Nakamura:2010zzi}\\
$m_t^{\rm pole}$                & $172.9$   & $1.1$   & $0$ & Gaussian &  \cite{Nakamura:2010zzi}\\
$\m_b(m_b)^{\overline{MS}}$                         & $4.19$    & $0.12$  & $0$ & Gaussian &  \cite{Nakamura:2010zzi}\\
$\alpha_s(M_Z)^{\overline{MS}}$                     & $0.1184$  & $0.0006$& $0$ & Gaussian &  \cite{Nakamura:2010zzi}\\
\hline \hline
\end{tabular}
\caption{The experimental measurements that constrain the CMSSM's
  parameters and the Standard Model's nuisance parameters. Masses are
 in GeV. The numbers in parentheses in the list of LEP experimental
  measurements are weaker experimental bounds, which we use for some
 sparticle mass hierarchies. }
\label{tab:exp_constraints}
\end{center}
\end{table}

\section{LHC and DD constraints}\label{lhcdd:sec}

\subsection{The \bfseries $\alpha_T$ SUSY search}\label{aphat:sec}

In order to calculate the likelihood for the LHC signal, we generally
follow the procedure detailed in Ref.~\cite{Fowlie:2011mb}. With
respect to the previous paper, we extend the efficiency maps to the
larger values of $m_0$ and $m_{1/2}$ considered here, and improve the
evaluation of the detector efficiency. While in~\cite{Fowlie:2011mb}
the geometrical acceptance was applied to simulate the CMS
detector response, here we apply the kinematical cuts detailed in the
experimental paper~\cite{Chatrchyan:2011zy} with the fast detector
simulator PGS4~\cite{PGS4}, whose parameters have been set to the
specifications given by the CMS Collaboration. Besides, we slightly
improve the minimization procedure for the difference in missing
transverse energy of the pseudojets. We have compared our efficiency
maps for $A_0=0$ and $\tan\beta=10$ with the results
of~\cite{Fowlie:2011mb} in the overlapping region, finding excellent
agreement.

\subsection{XENON100 bound}\label{x100:sec}

We incorporate in our scan the constraints induced on the SI
neutralino-proton cross section $\sigma_p^{\textrm{SI}}$ by the 90\%
C.L. upper bound $\sigma_{p,90}^{\textrm{SI}}$, which was recently
published by the XENON100
Collaboration~\cite{arXiv:1107.2155,Aprile:2011hx,arXiv:1104.2549}. As
described in the experimental papers, the bound is obtained from the
$p$-value of a likelihood function which includes the systematic and
statistical error in signal and background, as well as the
uncertainties on scintillation efficiency and escape velocity;
see~\cite{Aprile:2011hx} for details. However, the experimental
analysis does not consider the astrophysical uncertainties associated
with the chosen velocity distribution and DM halo
profile~\cite{Green:2010gw,Pato:2010zk,Arina:2011si}, nor the nuclear
physics uncertainties associated with calculations of the
$\pi$-nucleon sigma term $\Sigma_{\pi
  N}$~\cite{arXiv:0801.3656}, which are in fact
dominant. Here and below we denote these theoretical uncertainties
with $\tau$. (In contrast, we label the experimental uncertainties
with $\sigma$.) The uncertainties $\tau$ have been quantified in the
literature and amount up to approximately 10 times the upper bound
on the cross section, $\tau\sim
10\times\sigma_{p,90}^{\textrm{SI}}$~\cite{Fowlie:2011mb}. We
include $\tau$ in the likelihood, following the procedure described
in~\cite{deAustri:2006pe} for the treatment of upper bounds. This
procedure allows for a systematic treatment of the 
mismatch between true ($\hat\xi$) and approximate ($\xi$) theoretical
quantities; see~\cite{deAustri:2006pe} for details. In the limit of
negligible $\sigma$, the likelihood function is given by a convolution
of a step function, $\lim_{\sigma\rightarrow
  0}p(\sigma,\sigma_{p,90}^{\textrm{SI}}|\hat{\xi})\equiv\Theta(\sigma_{p,90}^{\textrm{SI}}-\hat{\xi})$,
and a Gaussian function parametrizing the theoretical uncertainties,
$p(\hat{\xi}|\xi,\tau)\equiv$Gauss$(\xi,\tau)$, where in this case
$\xi=\sigma_p^{\textrm{SI}}$. The result is a complementary
  error function likelihood which smears out the experimental bound
at the desired level,
\begin{equation}
p(\sigma_{p,90}^{\textrm{SI}}|\xi,\tau)=\frac{1}{2}\textrm{erfc}\left(\frac{\xi-\sigma_{p,90}^{\textrm{SI}}}
{\sqrt{2}\tau}\right)\,,\label{smearedbound}
\end{equation}
where the error function erfc($x$) is defined in the standard way,
\begin{equation}
\textrm{erfc}(x)\equiv\frac{2}{\sqrt{\pi}}\int_x^{\infty}e^{-t^2}dt\,.\label{erfc}
\end{equation}

Since it is obvious (and has been shown in~\cite{Fowlie:2011mb})
that such a large theoretical uncertainty greatly reduces the impact of the
XENON100 upper bound, for comparison we consider here the idealized case where $\tau$
is much reduced and is assumed to be comparable in size to
$\sigma_{p,90}^{\textrm{SI}}$. We do this to quantify the ``bare"
effect of the experimental bound on our PS.

Two recent papers~\cite{Bertone:2011nj,*Strege:2011pk} also considered
the impact of the XENON100 bound on CMSSM PS in the Bayesian
framework but their treatment of the theoretical uncertainties differs
from ours. Unlike us, they do not smear out the experimental bound,
but instead include the theoretical uncertainties 
in the scan as nuisance parameters and eventually marginalize over
them. While in principle this is an appropriate procedure in the contest of
Bayesian statistics, when uncertainties arise from
having a number of calculations which are based on different
assumptions and methodologies and thus yield incompatible
results, as is the case with the spread in the $\Sigma_{\pi N}$
uncertainty~\cite{arXiv:0801.3656}, it is difficult to believe that
they follow any statistical distribution. 

More importantly, as we will see below, even by choosing an extremely
optimistic (and, for the foreseeable future, rather unrealistically) small
value of $\tau$, say $\tau=\sigma_{p,90}^{\textrm{SI}}$, the XENON100
bound affects the PS only weakly. This is because current LHC
limits have a much stronger impact on the CMSSM parameters~\cite{Fowlie:2011mb}.
As we will show below, unless one neglects the theoretical uncertainties
altogether, thus taking the experimental bound at face value as a
hard cut, the impact of the bound $\sigma_{p,90}^{\textrm{SI}}$ on the
FP region remains limited.

\section{ID constraints}\label{id:sec}

Besides direct searches at accelerators and underground DM detectors,
the third direction in the quest for DM discovery is to use detectors
to look for uncharacteristic signatures pointing to physics beyond the
SM in cosmic ray radiation. One method for ID of DM is to exploit
searches for $\gamma$-ray spectra characteristic of DM
annihilation~\cite{arXiv:1108.3546,arXiv:0902.1089,arXiv:1001.4836,arXiv:1001.4531}. Another
one is the search for high energy neutrinos from annihilation of
weakly interactive massive particles (WIMPs) in the Sun or (to minor
extent) Earth in neutrino telescopes like
IceCube~\cite{arXiv:1111.5188}.

\subsection{The FermiLAT $\gamma$-ray telescope}\label{fermilat:sec}

We will focus in this subsection on $\gamma$ rays, leaving neutrinos
to the next. The FermiLAT detector~\cite{arXiv:0902.1089} in the Fermi
satellite has been used in the past few years on several targets for
detection of $\gamma$ rays from DM annihilation. Both multiwavelength
and line searches have been
performed~\cite{arXiv:1001.4836}. Recently, data were published by the
FermiLAT Collaboration which improved significantly the previous
sensitivities to DM searches. The most luminous source is the Galactic
center (GC) in the Milky Way, but it is also subject to a higher
astrophysical background. Better upper limits on the DM annihilation
cross section were obtained from the $\gamma$-ray fluxes of a few
dwarf spheroidal satellite galaxies
(dSphs)~\cite{arXiv:0909.3300,arXiv:1001.4531,arXiv:1108.3546} of the
Milky Way. They are less luminous but completely dominated by DM, with
little presence of gases or stars.

The differential flux
of DM annihilation-induced $\gamma$ rays from a generic target satellite galaxy is given by

\begin{equation}
\Phi(E)=\frac{N_{\gamma}(E)}{8\pi}\frac{\langle\sigma
v\rangle_{\textrm{ann}}}{m_{\chi}^2}\int_{\Delta\Omega} \int_{l_-}^{l_+} 
\rho^2[l(\theta)]dl(\theta)d\Omega\,,\label{gammaflux2}
\end{equation}
where $N_{\gamma}(E)$ is the photon energy distribution per
annihilation, $\langle\sigma v\rangle_{\textrm{ann}}$ is the
velocity-averaged pair-annihilation cross section, and $m_{\chi}$ is
the WIMP mass. $l$ is the line-of-sight distance, which depends on the
opening angle $\theta$ from the center of the galaxy; the integration
limits read $l_{\pm}=D\cos\theta\pm\sqrt{r_t^2-D^2\sin^2\theta}$,
where $D$ is the distance to the target galaxy, and $r_t$ is the tidal
radius of the DM halo.  $\rho(l)$ is the DM density distribution in
the galaxy, and the angular integration is performed about a solid
angle $\Delta\Omega$ calculated by pointing the $z$ axis of the
reference frame in the direction of the galaxy.  The double integral
in Eq.~(\ref{gammaflux2}) is commonly referred to as the $J$-factor,
whose determination depends on the selected DM galactic profile, and
on uncertainties in the stellar velocity distribution in the
galaxy. In a recent publication~\cite{arXiv:1108.3546}, the FermiLAT
Collaboration improved their previous analysis on
dSphs~\cite{arXiv:1001.4531} by (a) including two more dSphs in their
original set of eight and (b) incorporating the velocity
uncertainties on the $J$-factor in a likelihood approach. The DM
density distribution was assumed to follow a cuspy
Navarro-Frenk-White~(NFW) profile~\cite{Navarro:1995iw}, and its
normalization and scale factor were obtained by comparisons with
stellar velocity observations through the likelihood approach. By
applying this technique, the collaboration derived a stronger
95\%~C.L. upper bounds on the annihilation cross sections,
$\langle\sigma v\rangle_{i,95}$ as a function of $m_{\chi}$, for
different final-state channels ($i=b\bar{b}$, $\mu^-\mu^+$,
$\tau^-\tau^+$, $W^-W^+$). The bounds are reported in Fig.~2 of
Ref.~\cite{arXiv:1108.3546}.

In this paper we investigate the effects of the limits $\langle\sigma
v\rangle_{i,95}$ on the CMSSM PS. In order to effectively constrain
our Bayesian credible regions, we need to be able to invert the
procedure that the experimental group used to translate an upper limit
on the DM-related photon flux into a bound on a cross section.  The
95\%~C.L. upper bound on the photon flux as a function of energy
$\Phi_{95}(E)$ was not reported in~\cite{arXiv:1108.3546}. The
collaboration obtained it through the energy-binned likelihood
function given in their Eq.~(1). The bound depends nontrivially on
the flux actually observed by the detector, the known background, and
the astrophysical uncertainties.

We can approximately reconstruct $\Phi_{95}$ from any of
  the given limits $\langle\sigma v\rangle_{i,95}$ through
Eq.~(\ref{gammaflux2}), since each limit was obtained under the single-final-state assumption. To illustrate the procedure, let us select one final-state
channel, say $b\bar{b}$. The photon spectrum from $b\bar{b}$ cascades,
$N_{\gamma}^{(b\bar{b})}(E)$, does not depend on the nature of the DM
particle. Thus, we can eliminate the energy dependence by integrating
over the FermiLAT spectral sensitivity range, from 0.2 $\gev$ to 100
\gev: $N_{b\bar{b}}\equiv\int N_{\gamma}^{(b\bar{b})}(E)dE$. Assuming
the branching ratio BR$_{b\bar{b}}=1$, the upper bound on the
annihilation cross section as a function of $m_{\chi}$ is related to
the upper bound on the photon flux through Eq.~(\ref{gammaflux2}):
$\langle\sigma v\rangle_{b\bar{b},95}=8\pi
m_{\chi}^2\overline{\Phi}_{95}/(N_{b\bar{b}}J)$, where $J$ denotes
the $J$-factor and $\overline{\Phi}_{95}\equiv\int\Phi_{95}(E)dE$.

In the MSSM neutralino pair annihilation proceeds through several
final states, both at the tree and loop level. From the above
discussion it is easy to see that the 95\%~C.L.  bound on the total
annihilation cross section $\langle\sigma v\rangle_{\textrm{ann},95}$
of the MSSM can be obtained from the equality,
\begin{equation}
\overline{\Phi}_{95}=\frac{N_{b\bar{b}}\langle\sigma v\rangle_{b\bar{b},95}\, J}{8\pi m_{\chi}^{2}}=\frac{(\sum_{i=1}^{29}\textrm{BR}_i
N_i)\langle\sigma v\rangle_{\textrm{ann},95}\, J}{8\pi
m_{\chi}^{2}}\,,
\label{gammabound1}
\end{equation}
from which one obtains
\begin{equation}
\langle\sigma v\rangle_{\textrm{ann},95}=\frac{\langle\sigma v\rangle_{b\bar{b},95}\cdot
N_{b\bar{b}}}{\sum_{i=1}^{29}\textrm{BR}_i
N_i}\,,\label{gammabound}
\end{equation}
where we sum the branching ratios BR$_i\equiv\langle\sigma
v\rangle_i/\langle\sigma v\rangle_{\textrm{ann}}$ over all 29
channels implemented in SuperBayeS. In order to reduce the
uncertainties, we performed this procedure three times, using the
$\langle\sigma v\rangle_{b\bar{b},95}$, $\langle\sigma
v\rangle_{\mu^-\mu^+,95}$, and $\langle\sigma
v\rangle_{\tau^-\tau^+,95}$ lines. The resulting likelihoods were
eventually multiplied together. We neglected the $WW$ line because the
channel opens up only at $m_{\chi}\sim m_{W}$, giving less
constraining power on our PS than the other channels.

It has been long known from studies of the impact of $\gamma$-ray fluxes from the GC (see e.g.~\cite{Roszkowski:2007id}) that the uncertainties in the halo profile play a significant role in the determination of the derived cross section bounds.    
As Eq.~(\ref{gammaflux2}) shows, the assumptions on the DM halo profile affect the $J$-factor, sometimes by orders of magnitude, thus introducing an uncertainty in the photon flux received at the detector. 
On the other hand, a recent study on dSphs' DM density profiles~\cite{arXiv:1104.0412} showed that the analysis of stellar velocity dispersions of some dSph candidates imposes strong constraints on the mass within the half-light radius. 
When these constraints are taken into account, calculating the $J$-factor under the assumption of a NFW-like DM density profile yields photon fluxes which are comparable or somewhat lower than those obtained with smoother, corelike profiles. The cross sections given in~\cite{arXiv:1108.3546}  are therefore conservative, and the $1\sigma$ uncertainties resulting from the likelihood procedure amount
to approximately 3\%. 
In light of this, in our scan we considered the theoretical uncertainty to be zero and smeared
the 95\%~C.L. bound by a half Gaussian with uncertainty 3\%. This is
the Bayesian equivalent to performing a hard cut on the PS and
results from taking the limit $\tau\rightarrow 0$ in Eq.~(3.5) of
Ref.~\cite{deAustri:2006pe}.

\subsection{Neutrino telescopes}\label{neutrinotel:sec}


In 2010 the construction of all 86 strings of the IceCube neutrino
telescope at the South Pole was eventually
completed~\cite{Gaisser:2011iz}. The instrumented volume covers
approximately 1~km$^3$ and deploys 80 strings uniformly spaced 125~m
from one another, which carry 60 optical modules each. The energy
threshold of the muon effective area is estimated by the experimental
collaboration to be approximately
100~GeV~\cite{GonzalezGarcia:2009jc}. DeepCore, an additional array of
six denser strings, is also operational.  Its performance is
parametrized by the muon effective volume $V_{\textrm{eff}}$. The
energy dependence of $V_{\textrm{eff}}$ is plotted
in~\cite{Barger:2010ng}, and its parametrization can be found
in~\cite{Barger:2011em}. The resulting energy threshold is
$E_{\textrm{min}}\sim35$~GeV.

The IceCube Collaboration published in 2009 the first results on a
search for neutrinos from neutralino annihilation in the Sun, with
104.3 live days of data taking in their 22-string
configuration~\cite{Abbasi:2009uz}.  No excess over the atmospheric
background was detected. Recently, the collaboration published a new
report with the results from a combined analysis of 812 days of
lifetime data from AMANDA-II in the years $2001-2006$, as well as 149 days in
$2008-09$ for the 40-string IceCube
detector~\cite{Collaboration:2011e}. The data again show no excess
over the background. On the other hand, this implies a more stringent
upper bound on the spin-dependent (SD) neutralino-proton cross
section, and thus the data are starting to bite into regions of SUSY PS
that were previously probed only by DD experiments. While presently
available data do not allow us to draw any definite conclusions, the
results for the full configuration (which should be published soon)
will have significant constraining power, as noted already in
Refs.~\cite{Trotta:2009gr,*Bertone:2011kb,Profumo:2011zj,Allahverdi:2009se,*Allahverdi:2012bi}. Moreover, in
recent years the study of the detector sensitivity has become more
elaborate, including increasingly detailed detector
simulations~\cite{Erkoca:2009by,Barger:2010ng,Barger:2011em}.

In this paper, we show the impact of the expected one-year
95\%~C.L. sensitivities for the full configuration of IceCube and
DeepCore on CMSSM PS, in light of the constraints from the LHC, DD,
and ID described in the previous sections. With respect
to~\cite{Trotta:2009gr,*Bertone:2011kb,Profumo:2011zj}, in addition to
a more accurate treatment of the most constraining limit from LHC
results, we update the effective areas and volumes to the most recent
values~\cite{Barger:2010ng,Barger:2011em} and investigate the
separate impacts of different detection channels at IceCube and
DeepCore. We briefly explain the formalism below.

The differential neutrino flux on Earth is given by
\begin{equation}
\frac{dN}{dE_{\nu}}=\frac{\Gamma_{\textrm{eq}}}{4\pi R_{\textrm{SE}}^2}\sum_i
\textrm{BR}_i \frac{dN_i}{dE_{\nu}}\,,\label{nuflux}
\end{equation}
where $\Gamma_{\textrm{eq}}$ is half of the WIMP capture rate, the
BR$_i$ are the branching ratios to the final products of neutralino
pair annihilation, and $R_{\textrm{SE}}$ is the Sun-Earth distance.
Upon arriving on Earth, the neutrinos undergo charged-current (CC)
interactions in and near the detector volume, thus producing a muon
flux $d\Phi_{\mu}/dE_{\mu}$, which can be detected by the optical
modules through Cherenkov radiation. In order to realistically
approximate the detector response, we follow the procedure given in
Appendix~B of Ref.~\cite{Barger:2010ng}.  The muon events observed by
the 80-string IceCube detector can either be labeled as
\textit{upward} (those that come from neutrinos undergoing
CC scattering outside the 1-km$^3$ instrumented volume) or as
\textit{contained}  (those for which the interaction happens
inside the detector).  We denote the differential flux of upward muons
detected by the optical modules with
$d\Phi_{\mu}/dE_{\mu}^{\textrm{f}}$, where the index ``f" is added to
distinguish it from the muon flux at the interaction point,
$d\Phi_{\mu}/dE_{\mu}$. The difference arises because of energy losses
due to ionization, bremsstrahlung, pair production and photonuclear
effects~\cite{Lipari:1991ut} that affect the spectrum of muons
transiting in ice. The upward event rates at the detector are given by
\begin{equation}
N_{\textrm{events}}^{\textrm{up}}=\int_{0}^{m_{\chi}}\frac{d\Phi_{\mu}}{dE_{\mu}^{\textrm{f}}}A_{\textrm{eff}}(E_{\mu}^{\textrm{f}})dE_{\mu}^{\textrm{f}}\,,\label{Nevents}
\end{equation}
where $A_{\textrm{eff}}(E_{\mu}^{\textrm{f}})$ is the muon effective
area parametrized in Ref.~\cite{GonzalezGarcia:2009jc}. We decided to
neglect the angular dependence of the effective area since, for the
signal, it integrates to a factor approximately equal to one. The
expected signal rate has to be compared with the atmospheric
background, which we adopted from Ref.~\cite{Barger:2010ng},
$N^{\textrm{up}}_{\textrm{BG}}\simeq 6.1$~yr$^{-1}$.

The contained rate, by definition, does not suffer from
propagation related losses and is given by
\begin{equation}
    N_{\textrm{events}}^{\textrm{C}}=\int_{E_{\textrm{thr}}}^{m_{\chi}}\frac{d\Phi_{\mu}}{dE_{\mu}}(1\textrm{
    km}^2)dE_{\mu}\,,
    \label{NeventsC}
\end{equation}
where the threshold $E_{\textrm{thr}}$ is conservatively fixed at 100~GeV.
The background is $N^{\textrm{C}}_{\textrm{BG}}\simeq 15.6$~yr$^{-1}$.

Notice that the upward and contained event rates and atmospheric
backgrounds are calculated for the time of the year the Sun spends
below the horizon, to reduce the background from down-going muons. So,
the rates are calculated for six months' time, between the March and
September equinoxes~\cite{Barger:2010ng,Barger:2011em}. For DeepCore,
the larger instrumented volume outside the detector will be used to
veto atmospheric muons from above the horizon, so the event rates are
calculated over a 12-month period.

The events rate at DeepCore is given by
\begin{equation}
    N_{\textrm{events}}^{\textrm{DC}}=\int_{E_{\textrm{min}}}^{m_{\chi}}\frac{1}{L}\frac{d\Phi_{\mu}}{dE_{\mu}}
    V_{\textrm{eff}}(E_{\mu})dE_{\mu}\,,\label{NeventsDC}
\end{equation}
where the typical scale of the detector is $L=1$~km, and
$V_{\textrm{eff}}(E_{\mu})$ is the effective volume shown
in~\cite{Barger:2010ng}. The background is $N^{\textrm{DC}}_{\textrm{BG}}\simeq 2.5$~yr$^{-1}$.

We calculated the signal using a modified version of SuperBayeS, where
we updated the muon effective area and effective volume to the values
cited above. Notice that SuperBayeS (following
DarkSusy~\cite{darksusy}) uses tables from
WimpSim~\cite{Blennow:2007tw} for the propagation of the neutrino flux
through the Sun and to Earth, so that matter effects and neutrino
oscillations are properly accounted for. Also, the signal is
calculated without the approximation of thermodynamical
equilibrium~\cite{darksusy}, so that the rates slightly depend on
$\langle\sigma v\rangle_{\textrm{ann}}$. We fixed the local DM density
to 0.3 GeV/cm$^3$, and the root-mean-square of the velocity dispersion
in the halo to 270 km/s.  We compared our spectra and rates with the
code used for the calculations performed in~\cite{Barger:2010ng} and
found very good agreement. Our 95\%~C.L. (or $2\sigma$) projected
sensitivities are calculated in the standard way, with
$N_{\textrm{events}}/\sqrt{N_{\textrm{BG}}}=2$.

\section{Results in the CMSSM}\label{results:sec}

In this section we will present our numerical results. First, to set
the ground, we will display regions of the CMSSM mass parameters
favored by current data from the usual non-LHC constraints, from the
$\alpha_T$ constraint at the LHC, as well as from the XENON100 limit
for which we will consider two limiting cases of the theoretical
error. Next, we will add the constraint from dwarf spheroidals.  Finally, we will discuss the projected sensitivities at
IceCube/DeepCore.

In all plots we denote as ``non-LHC'' the usual set of relevant
constraints summarized in the first part of
Table~\ref{tab:exp_constraints}, while the $\alpha_T$ bound is
indicated with ``LHC\,($\alpha_T$)." The constraints from DD
are labeled according to the size of the theoretical uncertainties
assumed for them (see Sec.~\ref{x100:sec}). In
particular, when the scan is performed with
the theoretical uncertainty $\tau=10\times\sigma_{p,90}^{\textrm{SI}}$,
we label this with ``X100($\tau$=10)"; and when we apply
$\tau=1\times\sigma_{p,90}^{\textrm{SI}}$, we use the label
``X100\,($\tau$=1)." For the dSphs bound, instead, the assumption of a NFW DM halo profile is justified by recent studies, as explained in Sec.~\ref{id:sec}. We label the ID set of constraints with ``dSphs\,(NFW)."

\subsection{Impact on the CMSSM parameter space}

\begin{figure}[ht!]
\centering
\label{OLDLHCdd_m0_m12}%
\includegraphics[width=80mm,
height=80mm]{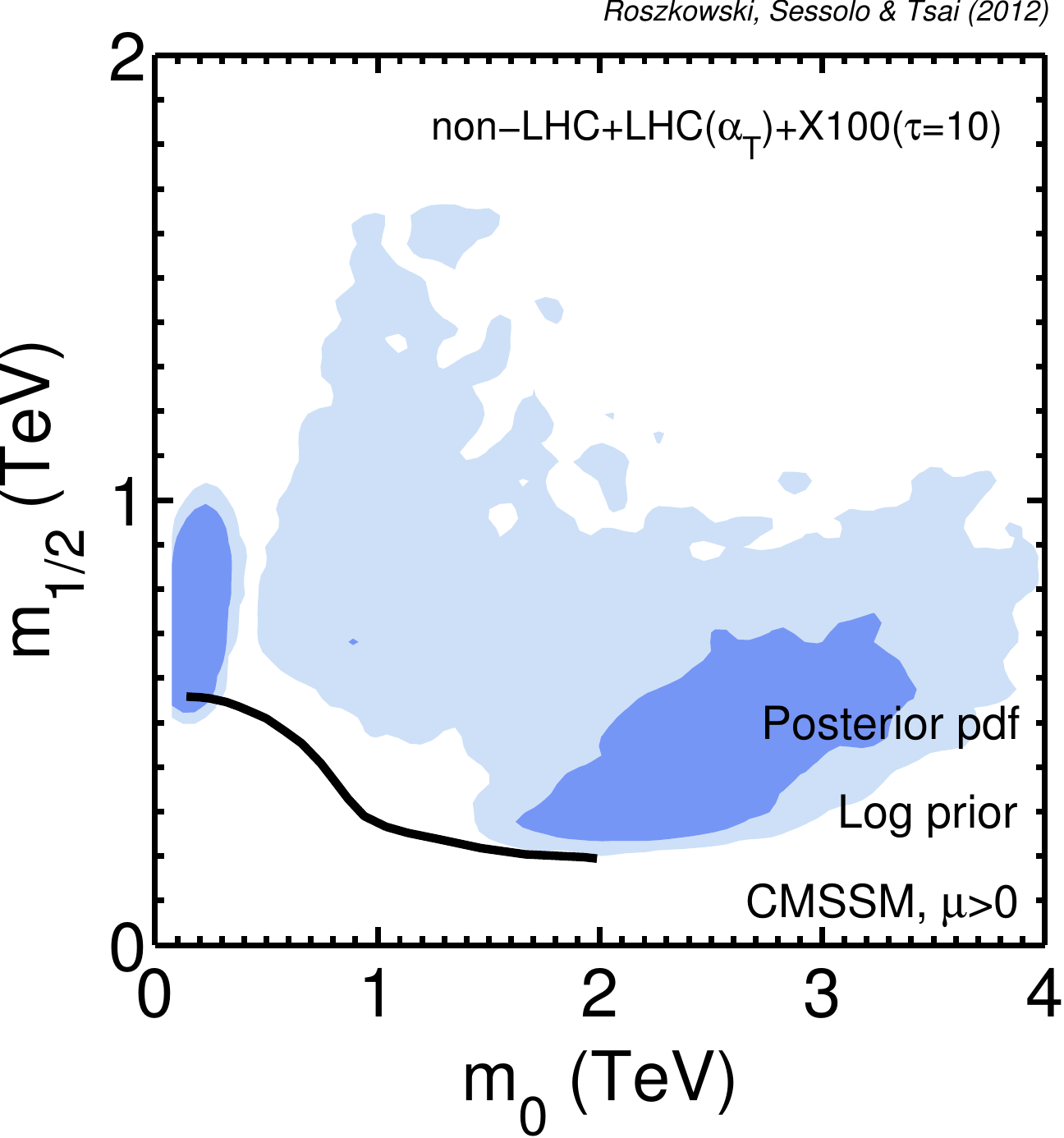}%
\label{OLDLHCddt1_m0_m12}%
\includegraphics[width=80mm,
height=80mm]{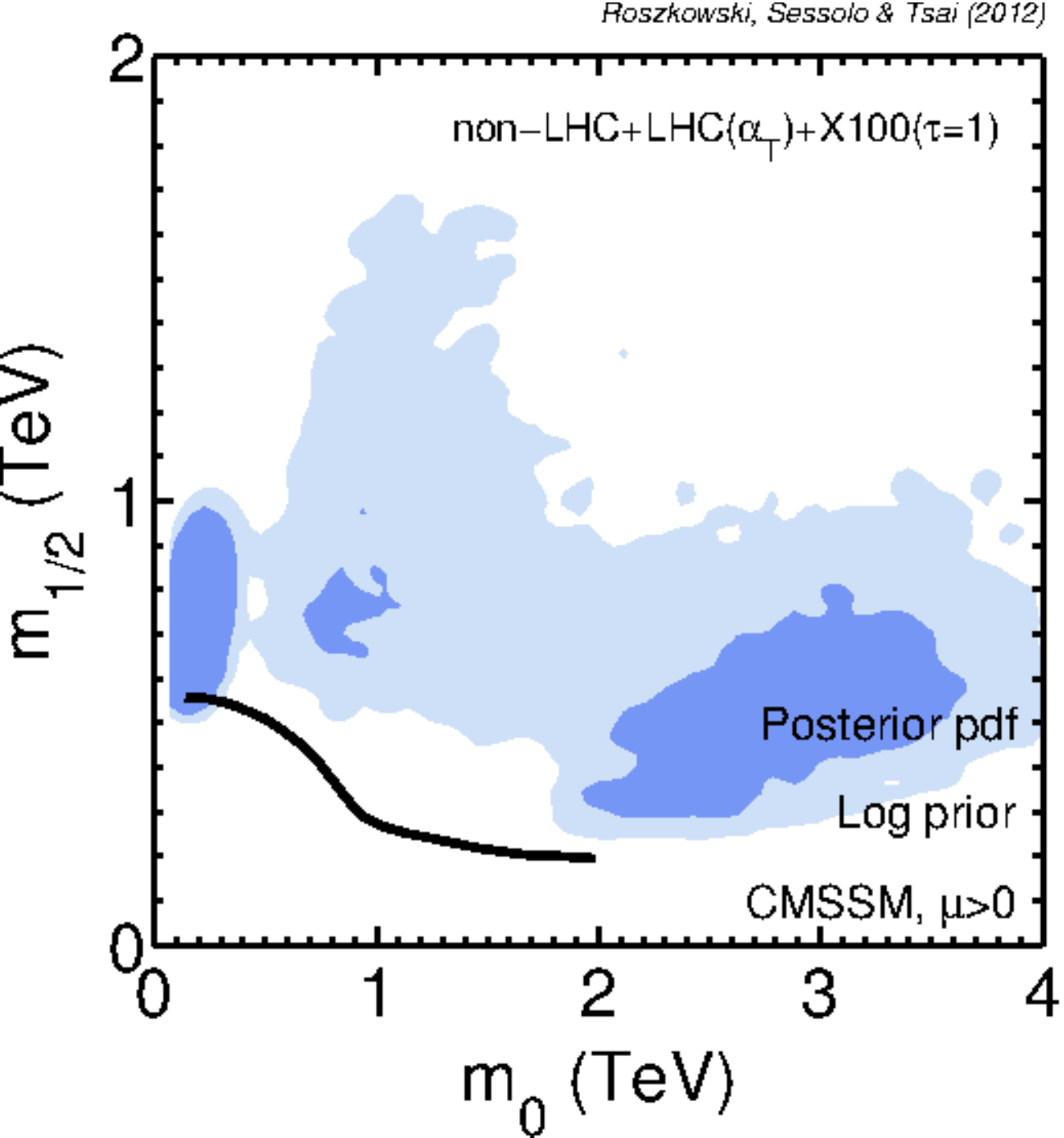}\\
\caption{Marginalized posterior pdf in the ($m_0,m_{1/2}$) plane. The
  68\%~C.L. credible regions are shown in dark blue and the
  95\%~C.L. regions in light blue. (a) The CMSSM parameters
  constrained by the non-LHC, $\alpha_T$, and XENON100
  data. (b) The same constraints as in (a), but the
  theoretical uncertainty on XENON100 is strongly reduced ($\tau=1$). The solid
  black line shows the 95\%~C.L. exclusion bound from the $\alpha_T$
  analysis of CMS.} 
\end{figure}

To start with, in Fig.~1(a) we show the marginalized
posterior pdf in the ($m_0$, $m_{1/2}$) plane for the set of non-LHC
constraints, the LHC $\alpha_T$ constraint from the CMS analysis of
1.1~fb$^{-1}$ data, and the XENON100 90\%~C.L. bound on
$\sigma_p^{\textrm{SI}}$ with a conservative estimate of the theory
error $\tau=10\times\sigma_{p,90}^{\textrm{SI}}$. Our intention here
is to summarize the present impact of experimental limits on the PS of
the CMSSM before we move to including the ID observations and
sensitivities. We show the 68\%~credible posterior regions in dark blue
and the 95\%~credible regions in light blue. The black line indicates the
CMS 95\%~C.L. exclusion limit~\cite{Chatrchyan:2011zy}, although we
stress that, as previously in~\cite{Fowlie:2011mb}, we apply it
through our own approximate likelihood function, as discussed in
Sec.~\ref{lhcdd:sec}.

The relative impact of the constraints applied in
Fig.~1(a) has been extensively analyzed
in~\cite{Fowlie:2011mb}, for a smaller range of $m_0$ and
$m_{1/2}$. In particular, one can see the two distinct 68\% posterior credible regions already found in~\cite{Fowlie:2011mb}. One of
these peaks in the stau-coannihilation 
region of the CMSSM, where $m_0$ is relatively small and
$m_{1/2}\gsim500$~GeV. (The $A$-resonance funnel actually extends to the much larger
$2\sigma$ region.) The other
appears in the FP/HB region at large $m_0$ and extends to much larger
values of $m_0$ than those considered in~\cite{Fowlie:2011mb}. As
noted in~\cite{Fowlie:2011mb}, the effect of the $\alpha_T$ bound is
mostly to push the high-probability credible regions up and outside
the experimental exclusion contour.

Figure~1(a) includes the 90\%~C.L. bound on
$\sigma_p^{\textrm{SI}}$ with a conservative theoretical uncertainty,
$\tau=10\times\sigma_{p,90}^{\textrm{SI}}$, as explained in
Sec.~\ref{x100:sec}. We have checked that removing the bound from the
figure would make hardly any difference. The near independence of
Fig.~1(a) on the XENON100 limit reflects the fact
that at present the LHC limits are much stronger, while DD experiments
are still marred by large theoretical uncertainties. We thus confirm
and extend to larger parts of the FP/HB region the finding of
Ref.~\cite{Fowlie:2011mb} that the further effect of the current XENON100 bound on
the posterior is minimal.

In Fig.~1(b) we investigate the effect of assuming
a highly optimistic uncertainty of $\tau=1\times\sigma_{p,90}^{\textrm{SI}}$,
as described in Sec.~\ref{x100:sec}. One can see in
Fig.~1(b) that it affects the FP/HB region
by reducing the area of the 95\%~credible region above
$m_{1/2}\sim200$~GeV at large $m_0$. The reason for this is clear: in
the FP/HB region, the higgsino component of the neutralino increases,
and the SI cross section with nuclei is enhanced; it thus becomes more
constrained by DD limits. As a consequence of this stronger constraint on the FP/HB region, some probability is shifted towards the $A$-resonance funnel region, thus creating a 68\% credibility ``island" at $m_0\sim 1$~TeV, which now becomes favored by the constraints from $b$~physics and $(g-2)_{\mu}$. On the other hand, it appears that even by
reducing the theoretical uncertainties on DD to only 100\% of the
experimental upper bound (a highly unlikely possibility) the impact on
the PS is still not sizeable. Below we will perform a similar analysis
in the ($m_{\chi},\sigma_p^{\textrm{SI}}$) plane.

\begin{figure}[ht!]
\centering
\label{OLDLHCdddSphs_unc_log_m0_m12}%
\includegraphics[width=80mm,
height=80mm]{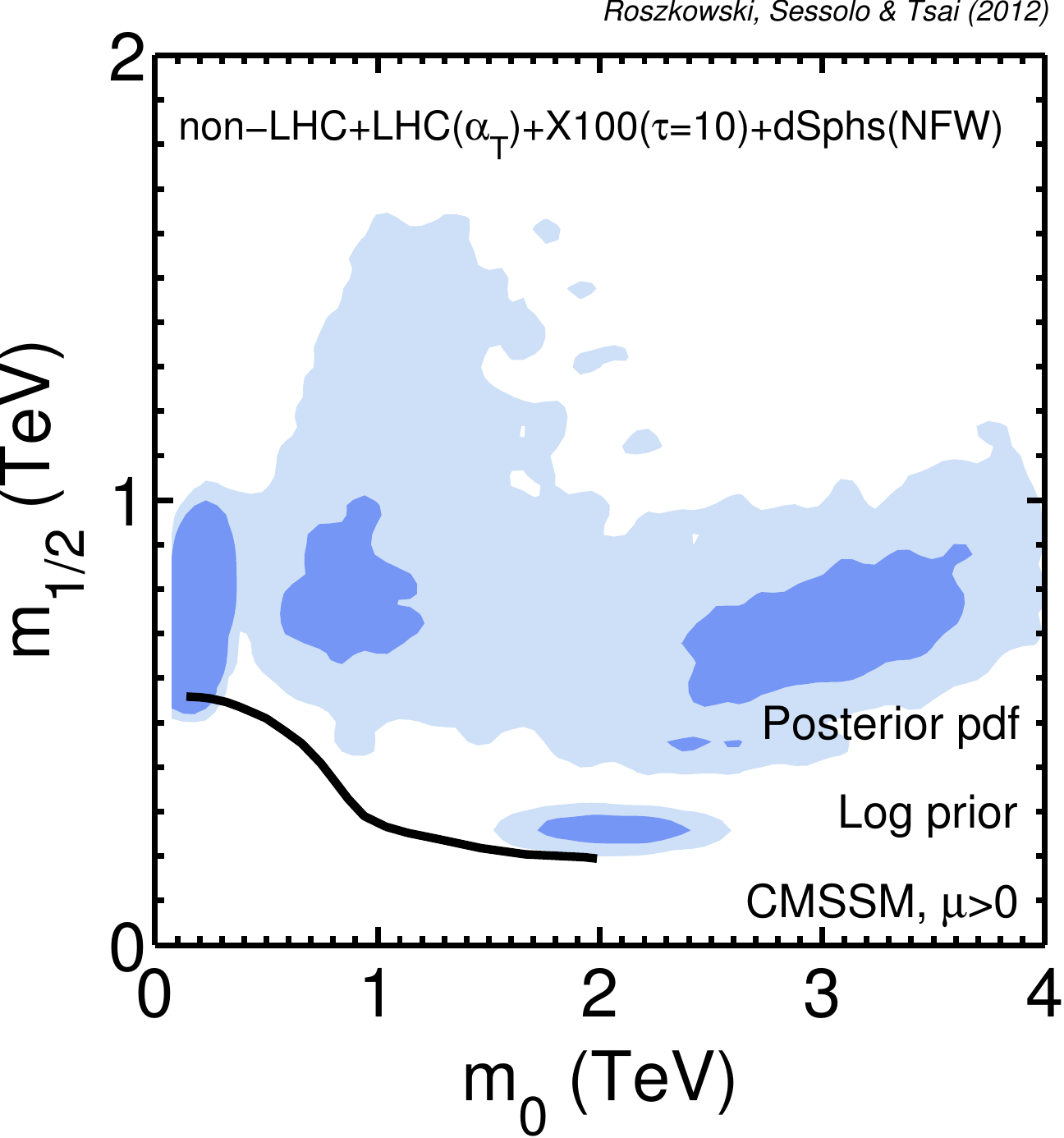}%
\label{OLDLHCdddSphs_nfw_log_m0_m12}%
\includegraphics[width=80mm,
height=80mm]{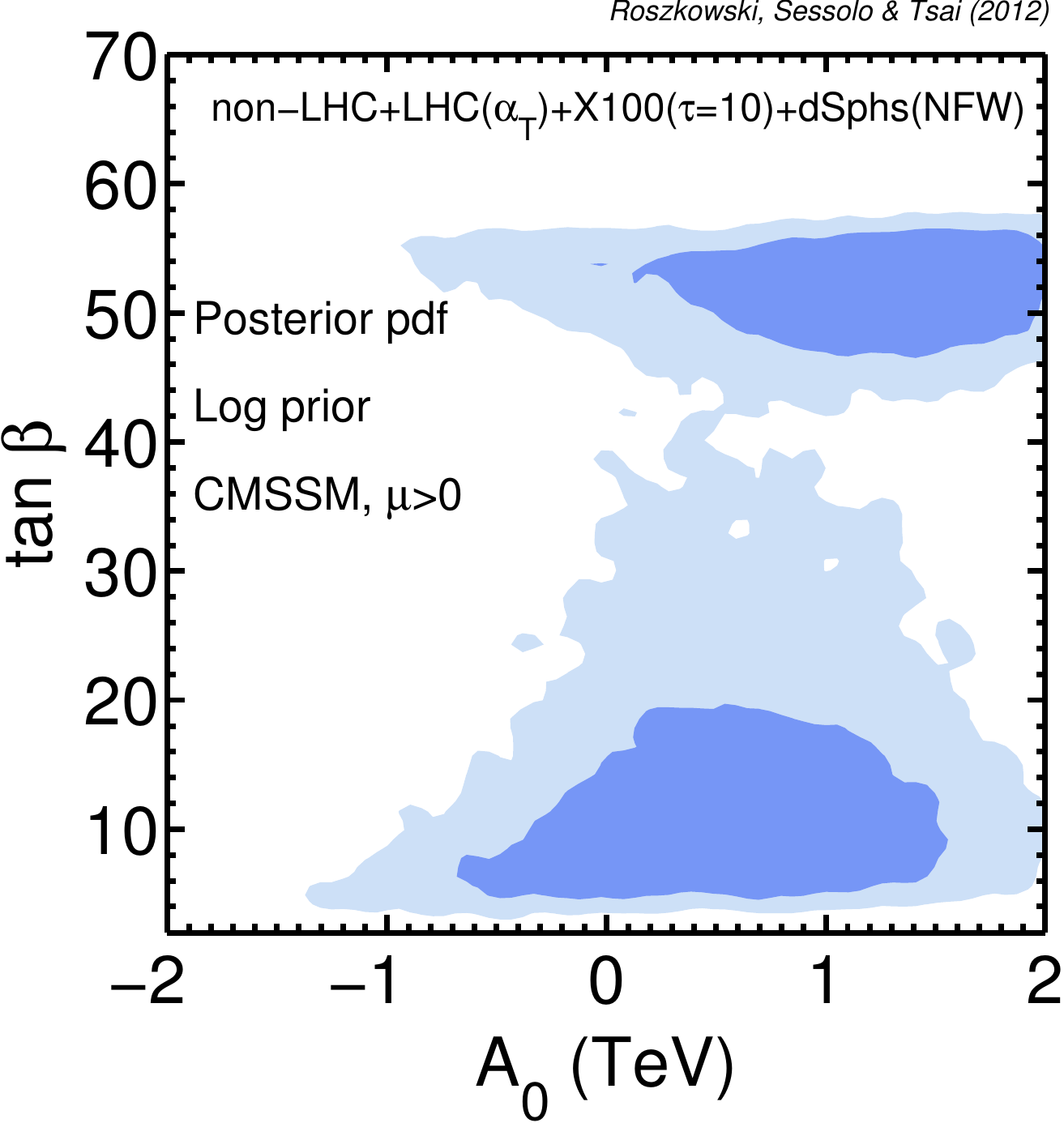}\\
\caption{(a) Marginalized posterior pdf in the ($m_0,m_{1/2}$) plane. The
  68\%~C.L. credible regions are shown in dark blue and the
  95\%~C.L. regions in light blue. The CMSSM parameters are
  constrained by the non-LHC, $\alpha_T$, XENON100, and dSphs
  data. The solid
  black line shows the $\alpha_T$ 95\%~C.L. exclusion
  bound. (b) Marginalized posterior pdf in the ($A_0,\tan\beta$) with the same constraints as in (a). }\label{OLDLHCdddSphs_log_m0_m12} 
\end{figure}

Next, in Figs.~2(a)$-$(b) we show the effects
of the dSphs bound on $\gamma$ rays from the FermiLAT Collaboration on
the marginalized posterior pdf in the ($m_0,m_{1/2}$) plane and the $(A_0,\tan\beta)$ plane, respectively. Figure~2(a) shows that the bound
from dSphs cuts significantly into the FP/HB region, and the impact is
stronger than observed with XENON100 alone, even for the idealized
case of extremely low theoretical uncertainties on $\sigma_p^{\textrm{SI}}$, depicted in
Fig.~1(b). One can see that the bound from dSphs
$\gamma$-ray fluxes significantly lowers the statistical relevance of
the FP region at $m_0\sim2500$~GeV, $m_{1/2}\sim400$~GeV. This is
because in this
region the enhanced higgsino component of the neutralino produces
larger $\langle\sigma v\rangle$, resulting in higher fluxes which in turn are more strongly constrained
by the FermiLAT data.

One can notice in Fig.~2(a) the appearance of a new high-probability region just above the $\alpha_T$ limit, at $m_0\sim2$~TeV and $m_{1/2}\sim300$~GeV. It is due to the Higgs-resonance mechanism through which the relic abundance constraint is satisfied, and it is not affected by the dSphs bound. Given the closeness of this island to the $\alpha_T$ upper bound, we investigated the possible effects on it of those LHC limits which are more sensitive to the region of PS at large $m_0$~\cite{CMS-PAS-SUS-11-008,CMS-PAS-SUS-11-004}. We applied a smeared cut at $m_{1/2}\sim280$~GeV (which, it is worth saying, is not a procedure that has the statistical rigor of our reconstructed $\alpha_T$ likelihood), and  we saw that the Higgs-resonance region disappeared. We conclude that this region is likely to vanish if we apply a more restrictive bound from the LHC. But we confirm that the main effects of the XENON100 and dSphs bounds on the rest of the PS is completely unchanged. In particular, the latter bound forces a shift in the high-probability areas of the FP/HB region towards higher values of $m_{1/2}$, as clearly seen by comparing Figs.~2(a) and 1(a).

The effect of the above constraints on the marginalized posterior pdf
in the $(A_0,\tan\beta)$ plane is presented in
Fig.~2(b). First, we have checked that,
without including the dSphs $\gamma$-ray constraint, our results
agreed well with those of Ref.~\cite{Fowlie:2011mb}.
However, since our scan spans a wider fraction of the FP/HB region, the region
around $\tan\beta\sim30-40$ gains higher statistical significance than
reported in that paper. When the dSphs constraint is not applied, it is contained in a 95\%~credible
region. This behavior is due to the much enlarged 1$\sigma$ credible
region at large $m_0$ and $m_{1/2}\gsim300$~GeV that can be observed
in Fig.~1(a). This region does not show up
in~\cite{Fowlie:2011mb} at the 1$\sigma$ level but only at the
2$\sigma$ level, and its presence here means that the
experimental constraints now favor a larger fraction of the FP/HB
region, which requires intermediate values of $\tan\beta$ to satisfy the relic abundance constraint.

On the other hand, one can see in
Fig.~2(a) that the dSphs
$\gamma$-ray flux constraint on the scattering and annihilation cross sections reduces the high-probability part of the FP region and also
pushes back the statistical significance of the region at
$\tan\beta\sim 40$ to less than 95\%~C.L. in Fig.~2(b).

\subsection{Effects on the observables}

We can better understand the effect of including the FP/HB region on
our scan by looking at the posterior distribution of the observables
relevant for DM detection: $\sigma_p^{\textrm{SI}}$,
$\langle\sigma v\rangle$, and the neutralino-proton SD cross section
$\sigma_p^{\textrm{SD}}$.

\begin{figure}[ht!]
\centering
\label{OLDLHCdd_mx_sigsip}%
\includegraphics[width=80mm,
height=80mm]{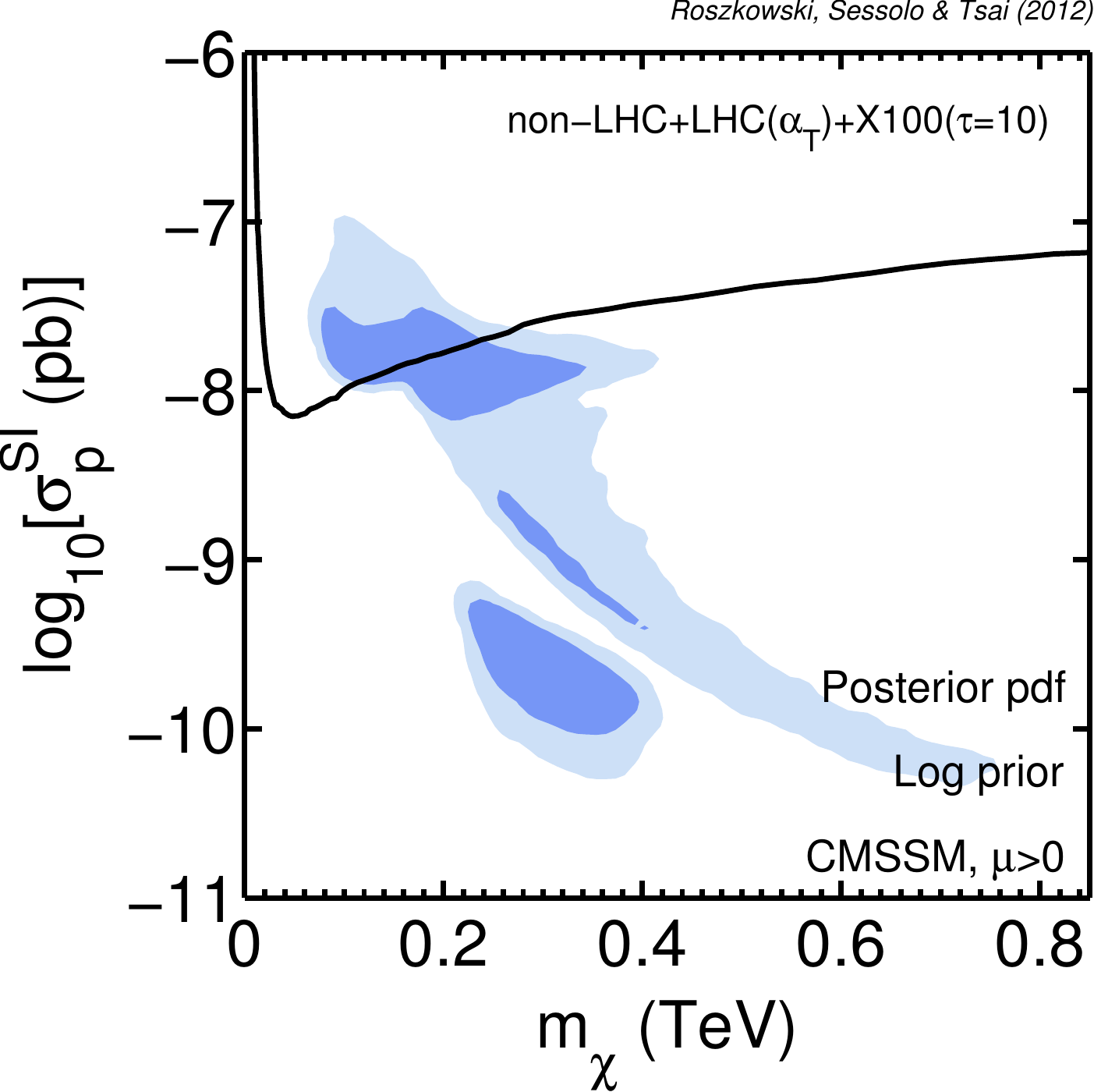}%
\label{OLDLHCdddSphs_nfw_log_mx_sigsip}%
\includegraphics[width=80mm,
height=80mm]{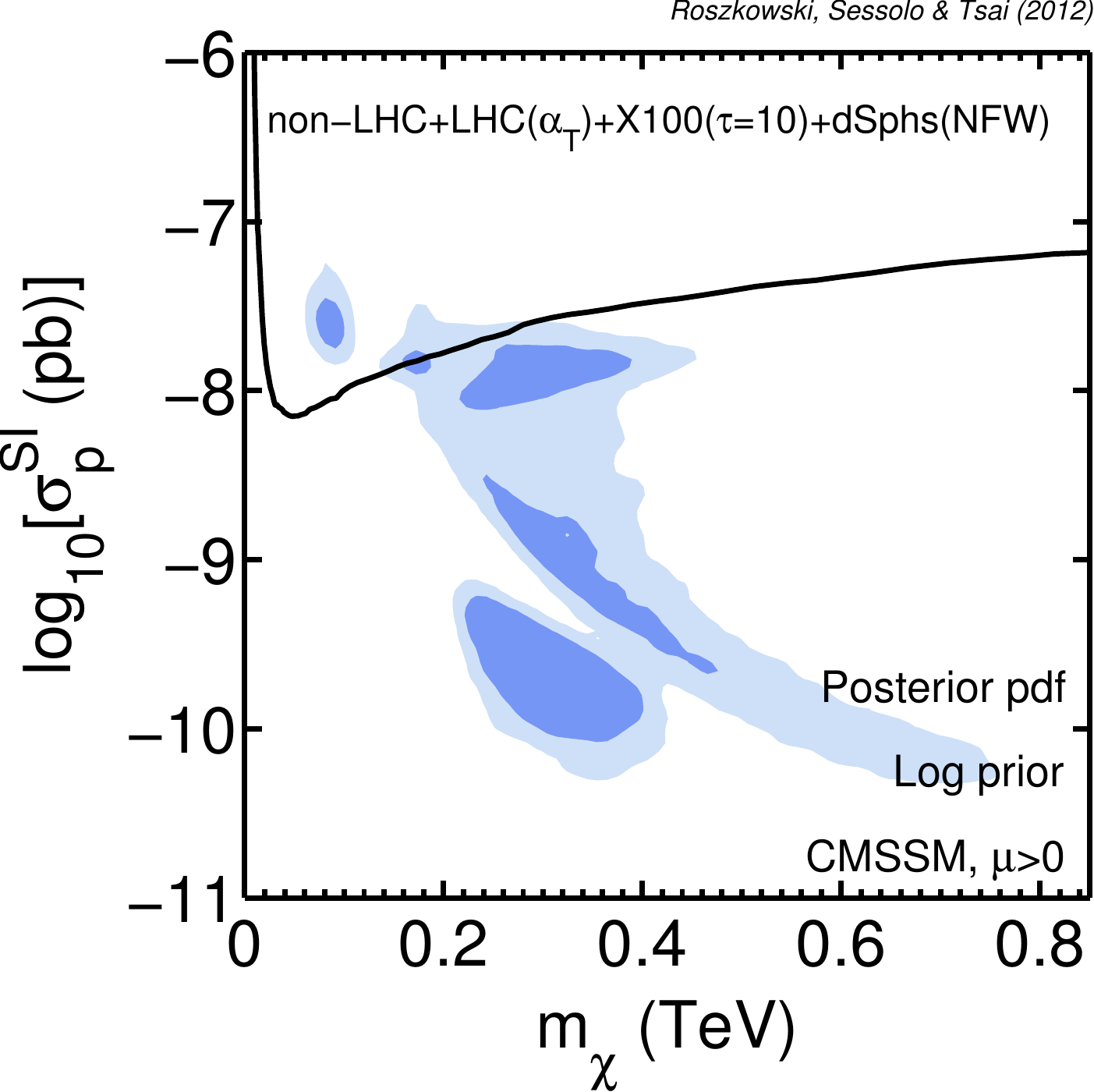}
\caption{Marginalized posterior pdf in the
  ($m_{\chi},\sigma_p^{\textrm{SI}}$) plane. The 68\%~C.L. credible regions
  are shown in dark blue and the 95\%~C.L. regions in light
  blue. (a) The impact of the non-LHC,
  $\alpha_T$, and XENON100 constraints. (b) The impact of the non-LHC, $\alpha_T$,
  XENON100, and dSphs constraints. The solid black line shows the XENON100
  90\%~C.L. exclusion bound.}\label{OLDLHCdd_log_mx_sigsip} 
\end{figure}

In Fig.~3(a) we show the posterior pdf projected
on the ($m_{\chi},\sigma_p^{\textrm{SI}}$) plane, with the non-LHC,
$\alpha_T$, and XENON100 (assuming
$\tau=10\times\sigma_{p,90}^{\textrm{SI}}$) constraints imposed. The
solid black line shows the XENON100 90\%~C.L. bound, although we
again emphasize that we implement it with a smearing through our
likelihood function and not as a hard cut. One can recognize three
familiar high-probability regions. A roughly horizontal one
corresponds to the FP/HB region. One (mostly 2$\sigma$ credible region
but with a narrow 1$\sigma$ piece) decreasing with $m_{\chi}$ is
given by the $A$-resonance effect. Finally, an island of high
probability at $200~\gev\lesssim\mchi\lesssim 400~\gev$ corresponds to
the stau-coannihilation region.

A comparison with the
corresponding figure in~\cite{Fowlie:2011mb} 
shows the
emergence of a sizeable 1$\sigma$ credible region above the
experimental bound, for lower $m_{\chi}\lsim 200$~GeV. As stated above,
this is a direct consequence of our extension to include a much larger
part of the FP/HB region. Figure~3(a) confirms
that the XENON100 bound itself does not much affect the statistical
significance of this region of PS, due to the large theoretical
uncertainties associated with $\sigma_p^{\textrm{SI}}$. 


The effect of applying the dSphs bound is shown in
Fig.~3(b). One can see a significant reduction of the posterior in the nearly horizontal FP/HB region at $\sigma_p^{\textrm{SI}}\sim 10^{-8}$~pb. 
One can also see that this reduction is comparable in size to those from the XENON100 bound if the theoretical uncertainties had a much reduced size. We have indeed checked that the case X100($\tau=1$) would not present noticeable differences with Fig.~3(b). This
fact was already hinted at when comparing
Figs.~1(b) and 2(a),
and shows how completely different experiments can in principle be competitive in
testing the CMSSM.
The small high-probability island that appears in the region at large $\sigma_p^{\textrm{SI}}$ and small $m_{\chi}$ of Fig.~3(b) corresponds to the Higgs-resonance region of the $(m_0,m_{1/2})$ plane, already seen in Fig.~2(a). We emphasize here that, in light of those LHC SUSY searches that are more sensitive to the large $m_0$ region, that island at small $m_{\chi}$ is likely to be excluded in future analyses. The overall reduction of the posterior in the FP/HB region, though, will be unaltered.

\begin{figure}[ht!]
\centering
\label{OLDLHC_mx_sigv}%
\includegraphics[width=80mm,
height=80mm]{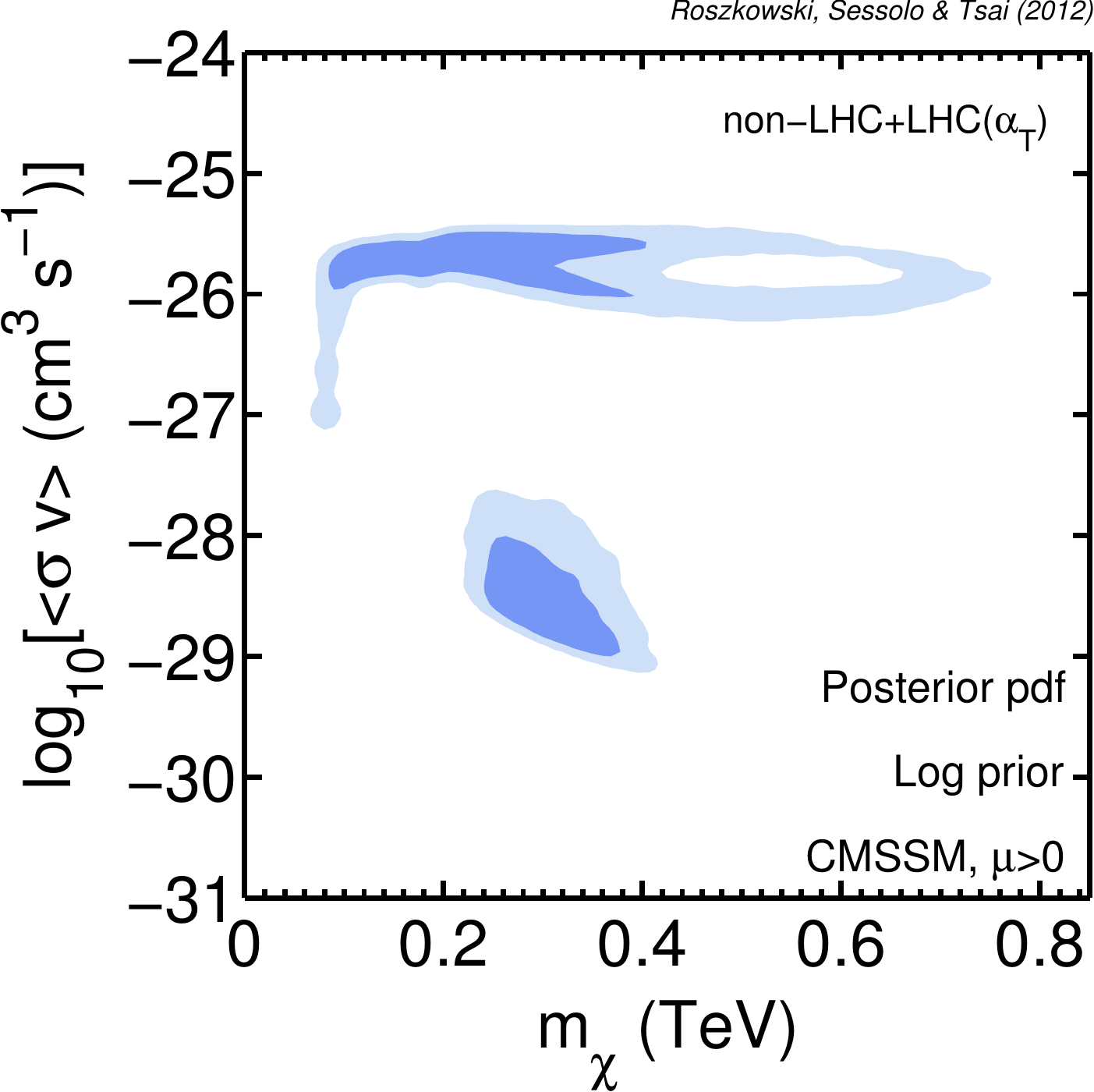}%
\label{OLDLHCdddSphs_nfw_log_mx_sigv}%
\includegraphics[width=80mm,
height=80mm]{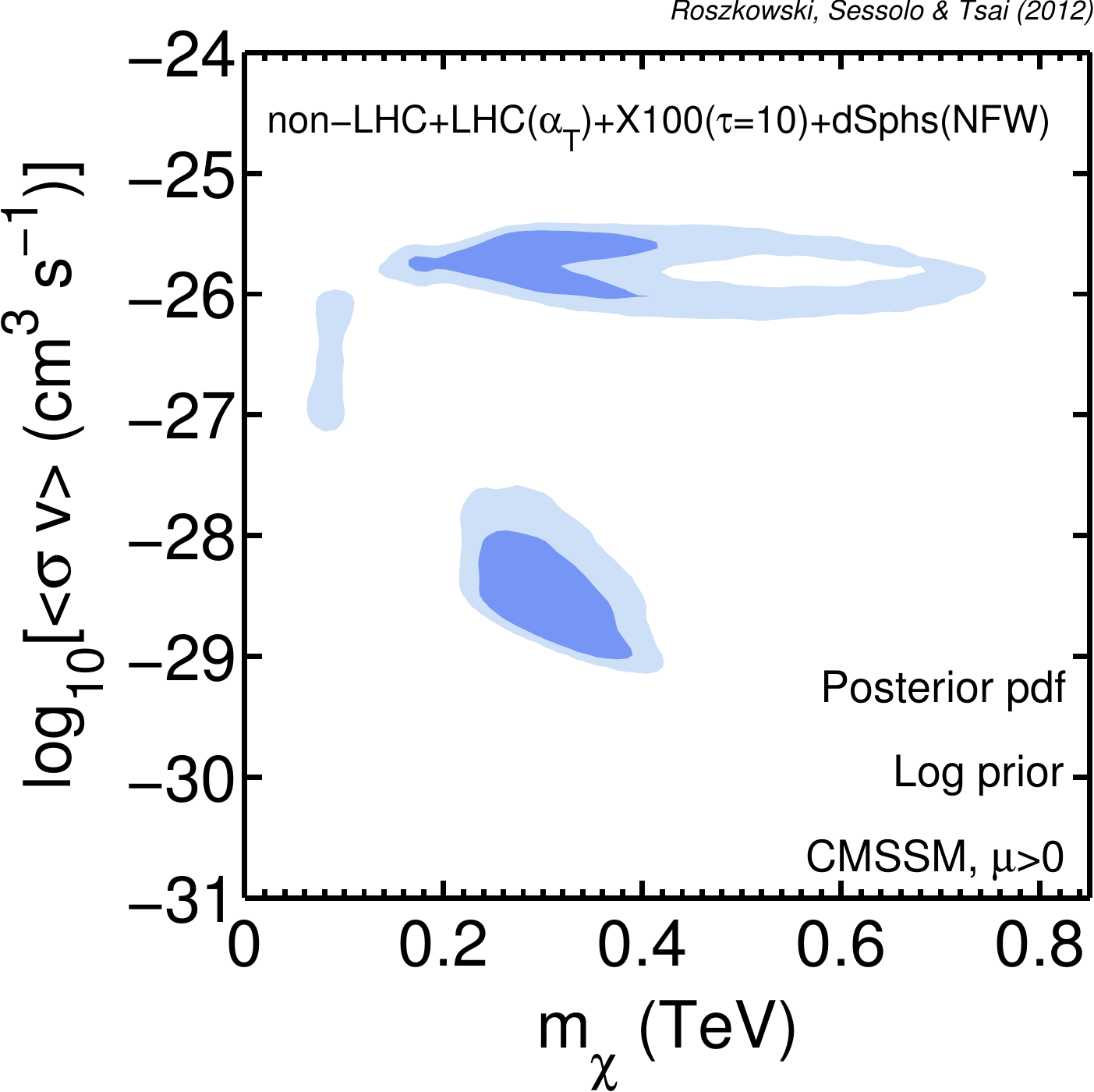}
\caption{\small Marginalized posterior pdf in the ($m_{\chi},\langle\sigma
  v\rangle_{\textrm{ann}}$) plane. The 68\%~C.L. credible regions are
  shown in dark blue and the 95\%~C.L. regions in light
  blue. (a) The impact of non-LHC and $\alpha_T$ constraints only. (b) The impact of non-LHC,
  $\alpha_T$, XENON100, and dSphs constraints.}\label{OLDLHCdddSphs_log_mx_sigv} 
\end{figure}

Figure~4 shows the marginalized
posterior pdf in the $(m_{\chi},\langle\sigma v\rangle)$ plane. To
start with, the effect of applying only the non-LHC and $\alpha_T$
constraints is presented in Fig.~4(a), where one
can see a familiar high-probability horizontal region at $\langle\sigma
v\rangle_{\textrm{ann}}\sim 3\times
10^{-26}$~cm$^{-3}$\,s$^{-1}$, which arises in order to satisfy the relic abundance
constraint from neutralino
pair-annihilation in the FP/HB and $A$-resonance regions. These two different mechanisms produce comparable but slightly different $\langle\sigma v\rangle_{\textrm{ann}}$, a fact that reflects in the appearance of a ``hole" in the posterior high-probability area. The hole is due to the conjunction of two different branches: the lower one, at $\langle\sigma
v\rangle_{\textrm{ann}}\sim
10^{-26}$~cm$^{-3}$\,s$^{-1}$, is due to annihilation in the $A$-funnel region of the CMSSM; the upper one, closer to $\langle\sigma
v\rangle_{\textrm{ann}}\sim
3\times 10^{-26}$~cm$^{-3}$\,s$^{-1}$, is due to annihilation in the FP/HB (and bulk) region instead.   
The little vertical ``tail" at $m_{\chi}\lesssim 100$~GeV is due to the Higgs-resonance region, and it would likely disappear if we considered the LHC searches most sensitive to the large $m_{0}$ region, as mentioned above.
Additionally, there is an island of high
probability at $200~\gev\lesssim\mchi\lesssim 400~\gev$ and much lower
$\langle\sigma v\rangle_{\textrm{ann}}$ corresponding to
the stau-coannihilation region, where the neutralino annihilation
cross section provides a subdominant contribution to the relic
density.

Figure~4(b)
shows the impact of dSphs when added to the non-LHC, $\alpha_T$, and
XENON100 constraints. 
It is clear that the dSphs bound disfavors the 1$\sigma$ credible region
at large $\langle\sigma v\rangle$ and 100~GeV$\lesssim m_{\chi}\lesssim 200$~GeV, as
over there one obtains the highest $\gamma$-ray flux through
the ratio $\langle\sigma v\rangle_{\textrm{ann}}/m_{\chi}^2$ in
Eq.~(\ref{gammaflux2}). On the other hand, the 1$\sigma$ credible
region at $m_{\chi}>200$~GeV remains unconstrained by DD and ID
experiments, even when idealized low uncertainties on DD are assumed.

\subsection{Effects due to IceCube}

\begin{figure}[ht!]
\centering
\label{LHCdddSphs_nfw_log_mx_sigsdp}%
\includegraphics[width=80mm,
height=80mm]{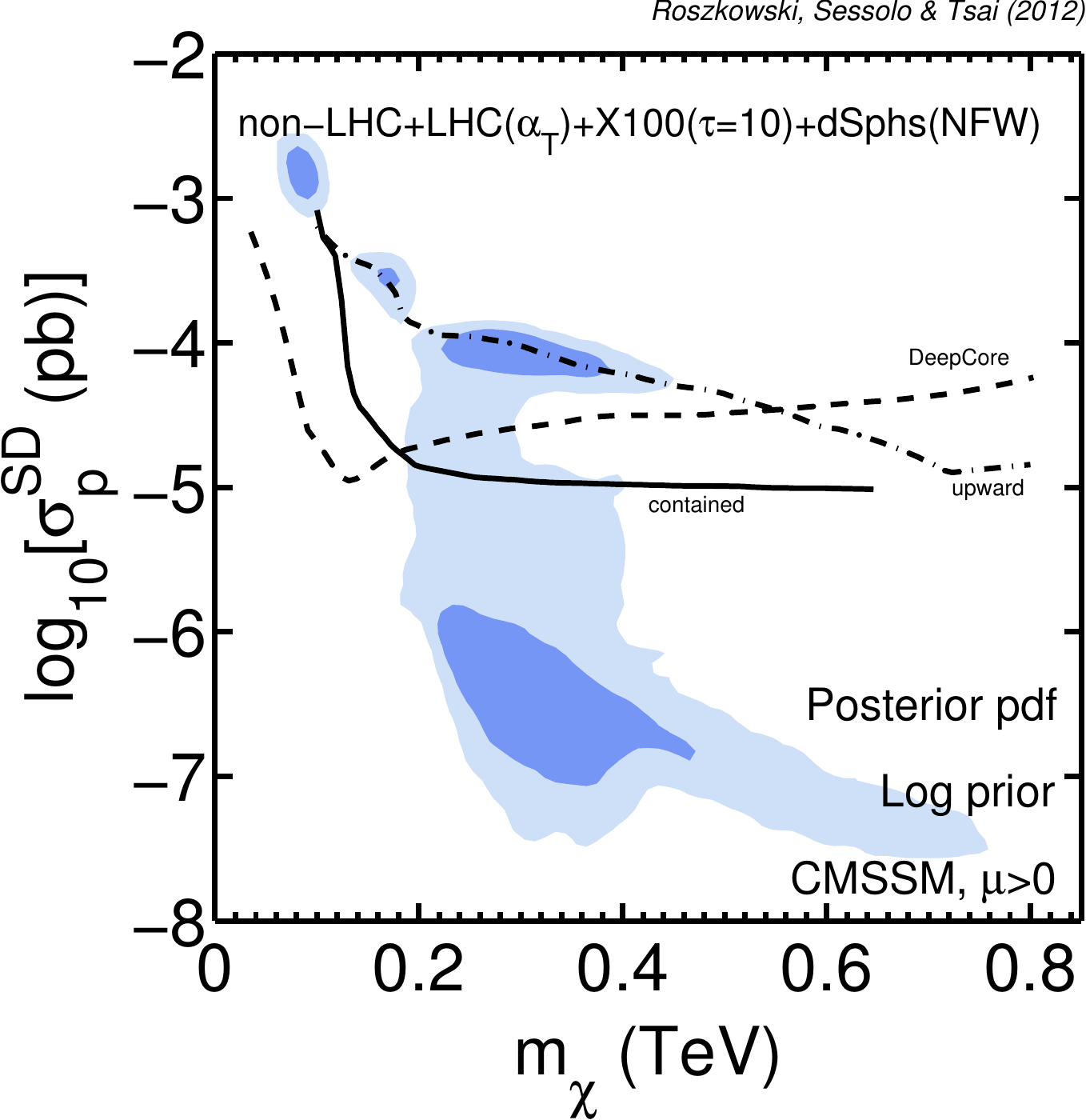}%
\label{OLDLHCddt1dSphs_nfw_log_mx_sigsdp}%
\includegraphics[width=80mm,
height=80mm]{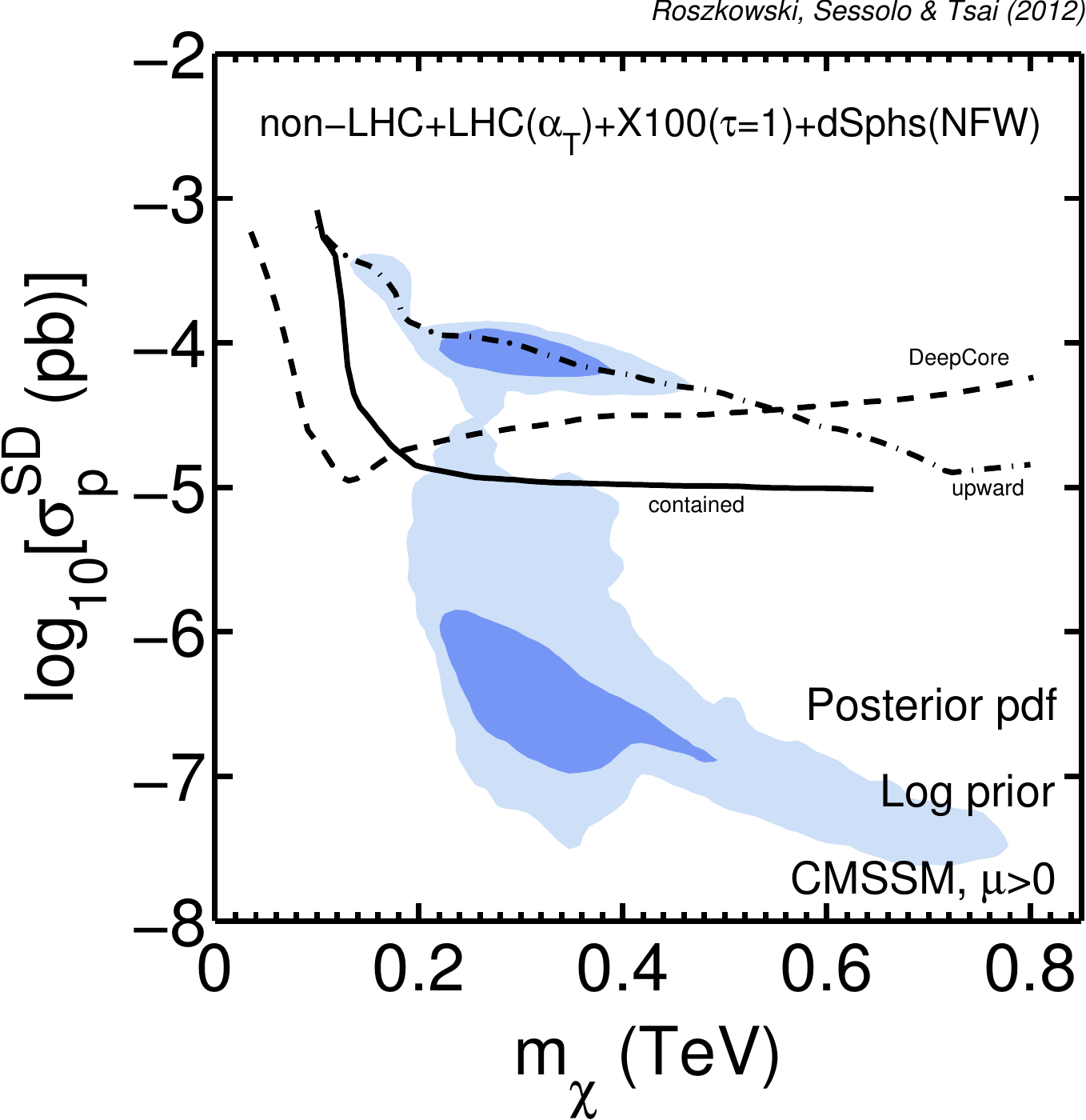}\\
\caption{Marginalized posterior pdf in the
  ($m_{\chi},\sigma_p^{\textrm{SD}}$) plane. The 68\%~C.L. credible
  regions are shown in dark blue and the 95\%~C.L. regions in light
  blue. (a) The impact of the non-LHC, $\alpha_T$, XENON100, and dSphs
  constraints.  (b) The impact of non-LHC,
  $\alpha_T$, XENON100, and dSphs constraints when the theoretical
  uncertainty on XENON100 is reduced. The
  solid line is the one-year 95\%~C.L. sensitivity in contained events
  at IceCube, the dashed line the corresponding sensitivity for DeepCore,
  and the dash-dotted line the sensitivity in upward
  events.}\label{OLDLHCddt1dSphs_log_mx_sigsdp} 
\end{figure}

In this subsection we analyze the effect on the CMSSM of the projected
one-year 95\%~C.L. experimental sensitivities for the 80-string
configuration at IceCube and the impact of the additional six
DeepCore strings. In the assumption of equilibrium in the Sun, the
event rates depend directly on the neutralino capture rate and,
consequently, are most strongly dependent on the SD scattering cross
section (although for some regions of PS the SI cross section can be
important~\cite{darksusy}).

In both panels of Fig.~5
we plot the marginalized pdf in the
$(m_{\chi},\sigma_p^{\textrm{SD}})$~plane, after applying all of the
constraints discussed above. In
Fig.~5(a) we show the case including the
large uncertainties for DD searches and in
Fig.~5(b) the one with reduced
uncertainties. The three black lines show the projected sensitivities
of the IceCube detection channels considered in this work. The
dash-dotted line shows the sensitivity for upward events at IceCube, the
solid line the sensitivity in contained events, and the dashed line
the sensitivity for DeepCore. We confirm the complementarity of the
three configurations to test different ranges in the neutralino
mass~\cite{Barger:2010ng,Barger:2011em}. Due to its denser array,
DeepCore has a lower energy threshold, so that it is better suited
to probe the region at low mass, $m_{\chi}\sim100$~GeV, where it could in principle compete with direct SUSY searches. As the DM mass
increases, though, the neutrino flux becomes suppressed by $1/m_{\chi}^2$, so
that the bigger volume available in the IceCube 80-string
configuration becomes more important. The result is increased sensitivities
in the larger detector, which can overcome the observational
capabilities of DeepCore.

\begin{figure}[ht!]
\centering
\label{OLDLHCdddSphs_nfw_log_mx_N_up}%
\includegraphics[width=70mm,
height=70mm]{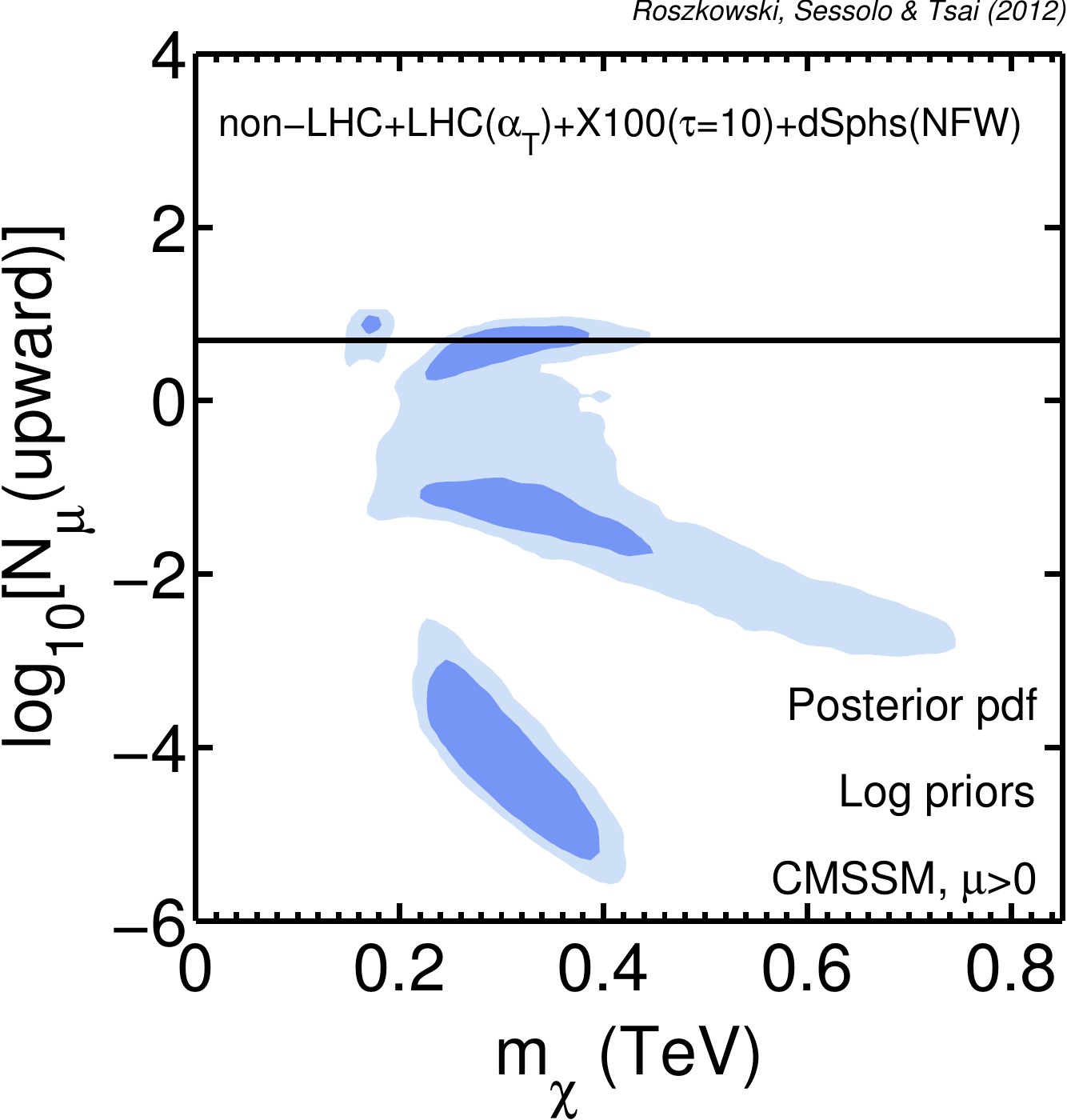}%
\label{OLDLHCdddSphs_nfw_log_mx_N_con}%
\includegraphics[width=70mm,
height=70mm]{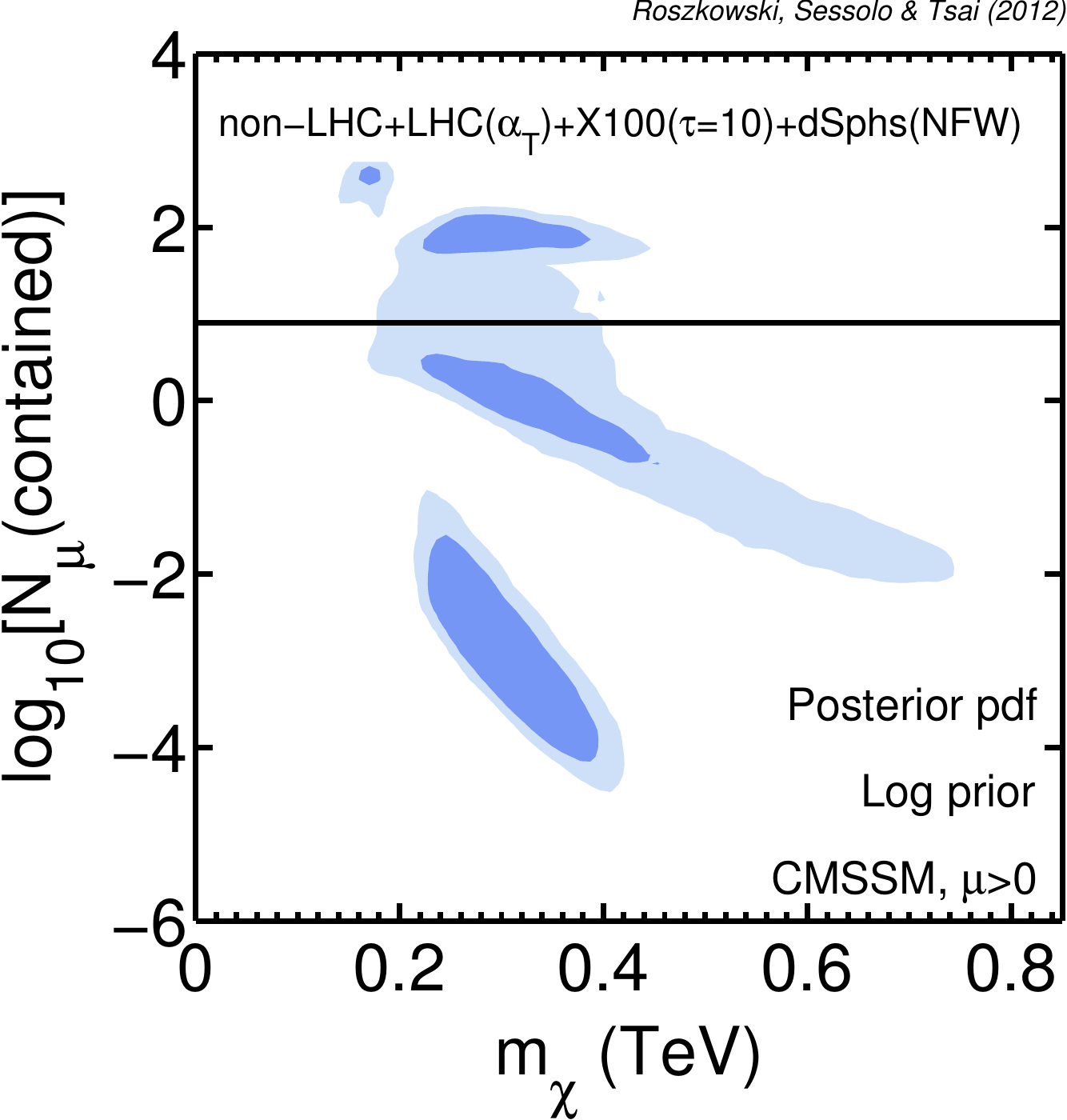}\\
\label{OLDLHCdddSphs_nfw_log_mx_N_DC}%
\includegraphics[width=70mm,
height=70mm]{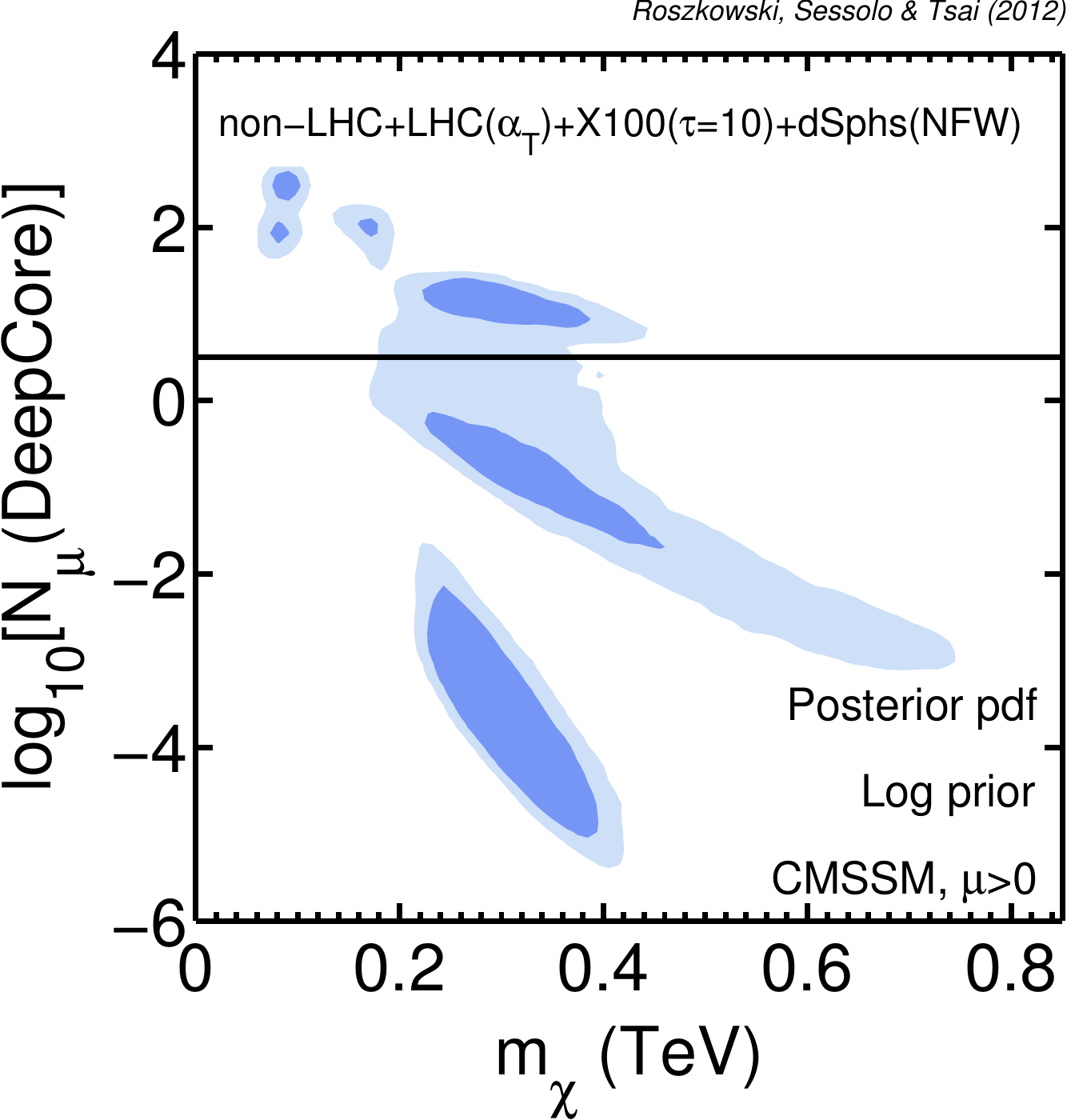}%
\label{OLDLHCdddSphs_nfw_log_m0_m12_DC}%
\includegraphics[width=70mm,
height=70mm]{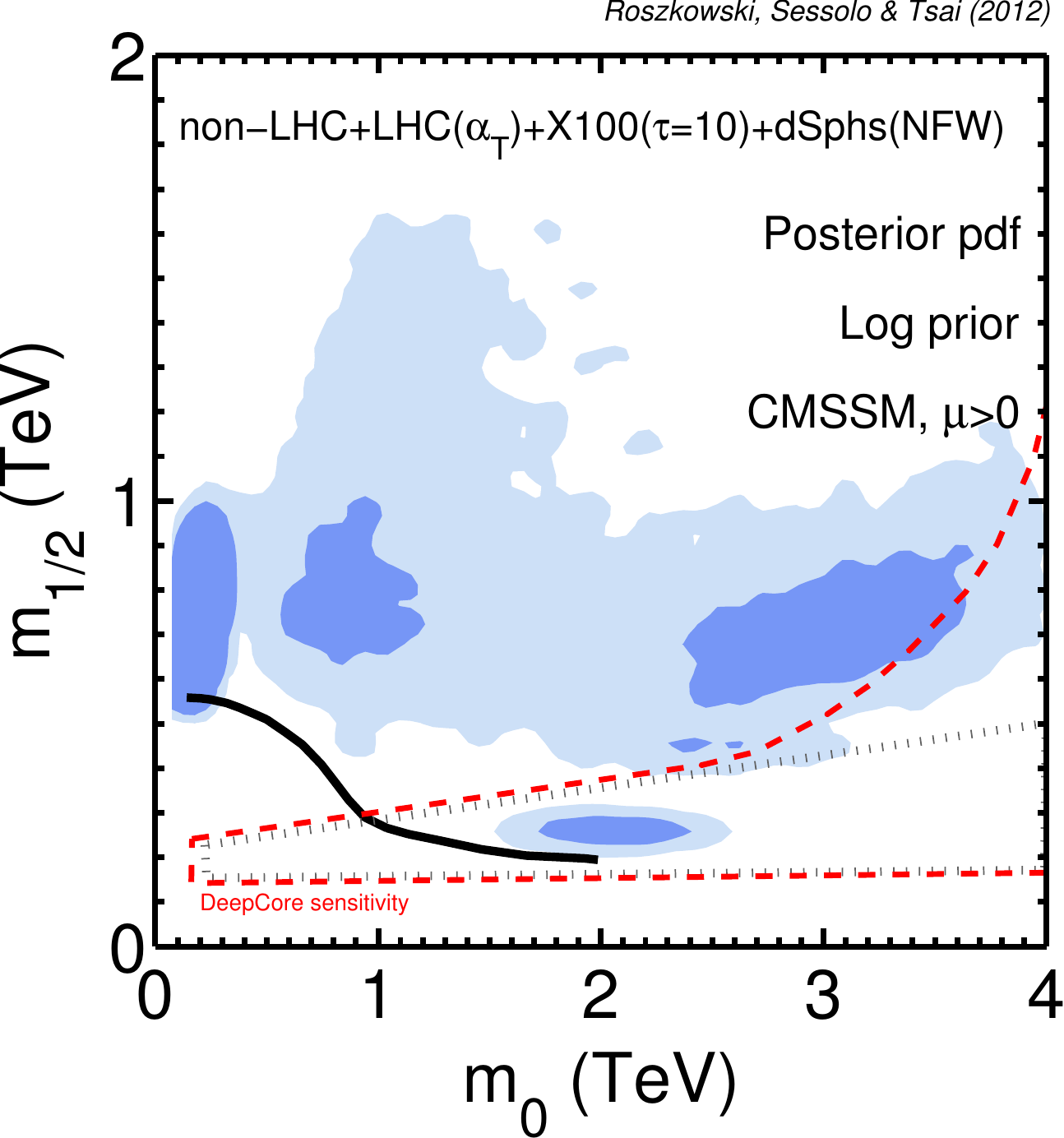}
\caption{(a)$-$(c) Marginalized posterior pdf in the
  ($m_{\chi},N_{\textrm{events}}$) plane. The 68\%~C.L. credible
  regions are shown in dark blue and the 95\%~C.L. regions in light
  blue. (d) Marginalized posterior pdf in the
  ($m_0,m_{1/2}$) plane with DeepCore one-year 95\%~C.L. sensitivity (inside red
  dashed lines, 90\% of the points are above sensitivity; inside gray dotted lines, 100\% of the points are above sensitivity). In all plots, the constraints are non-LHC, $\alpha_T$,
  XENON100, and dSphs. Upward events at
  IceCube are shown in (a), contained events in (b), and events at
  DeepCore in~(c).}\label{OLDLHCdddSphs_log_m0_m12_DC} 
\end{figure}

In the first three panels of Fig.~6, we
show the posterior probability in the
$(m_{\chi},N_{\textrm{events}})$~plane, where the muon rates at the
detector, $N_{\textrm{events}}^{\textrm{up}}$
[Fig.~6(a)],
$N_{\textrm{events}}^{\textrm{C}}$
[Fig.~6(b)], and
$N_{\textrm{events}}^{\textrm{DC}}$
[Fig.~6(c)] are given in
Eqs.~(\ref{Nevents}), (\ref{NeventsC}), and (\ref{NeventsDC}),
respectively. In dark blue one can see three 1$\sigma$ credible
regions in the mass range $200~\gev\lesssim\mchi\lesssim
400~\gev$. Since they correspond to the two high-probability regions
depicted in Fig.~5 for the same mass range, the splitting of the lower
regions indicates some dependence on SI interactions.

Additionally, in the low $m_{\chi}$ region of Fig.~6(c) one can spot three little high-probability islands. 
The one at the highest mass corresponds to the little 68\% credible region in the FP/HB portion of Fig.~6(d), at $m_{1/2}$ around $400-500$~GeV. 
The two connected islands at an even lower mass scale, instead, correspond to the Higgs-resonance region. 
Their separation might be an indication of slight sampling issues, due to the fact that there are not many scan points generated close to the $\alpha_T$ bound.

For contained events in IceCube, and for DeepCore, about $20-25\%$ of
the 2$\sigma$-favored PS is above the projected sensitivity, thus
indicating that the next round of data from the IceCube Collaboration
has the potential to significantly constrain the PS. In
Fig.~6(d) we show the region of PS in
the ($m_0,m_{1/2}$) plane which is above the sensitivity at
DeepCore. The region bounded by the gray (dotted) lines represents the region of parameter space for which \textit{all} of the points in the scan are above the DeepCore sensitivity. 
The region bounded by the red (dashed) lines yields instead approximately 90\% of the points above the DeepCore (or IceCube contained) sensitivity.
One can see that it extends beyond the region tested by the FermiLAT $\gamma$-ray flux from dSphs. 
It appears clear that, even if the Higgs-resonance region were to be excluded by higher LHC sensitivities in the large $m_0$ region, the impact of DeepCore/IceCube on the FP/HB region would nonetheless be significant.

\begin{figure}[ht!]
\centering
\label{contained_mx_events}%
\includegraphics[width=70mm,
height=70mm]{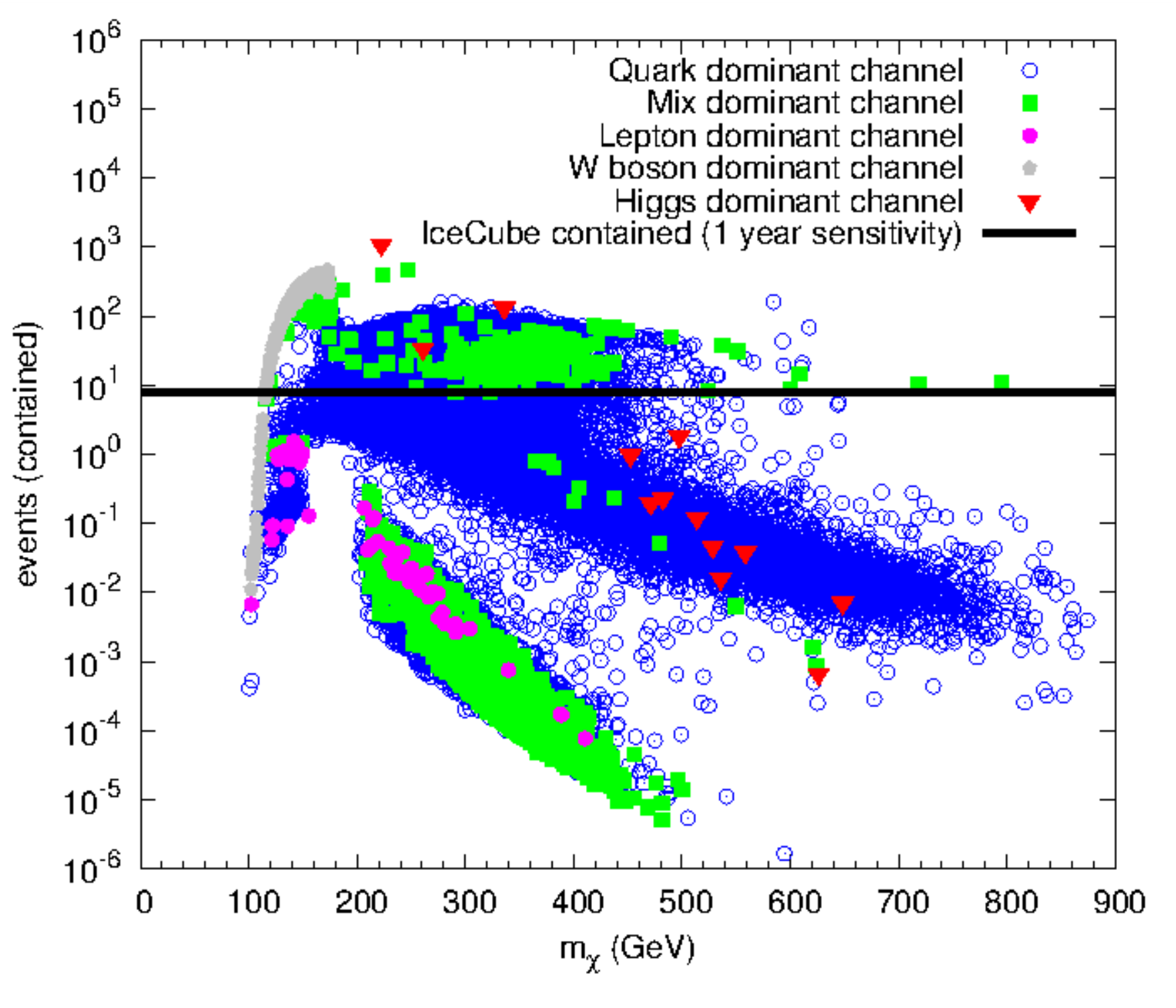}%
\label{Group_DeepCore_mx_events}%
\includegraphics[width=70mm,
height=70mm]{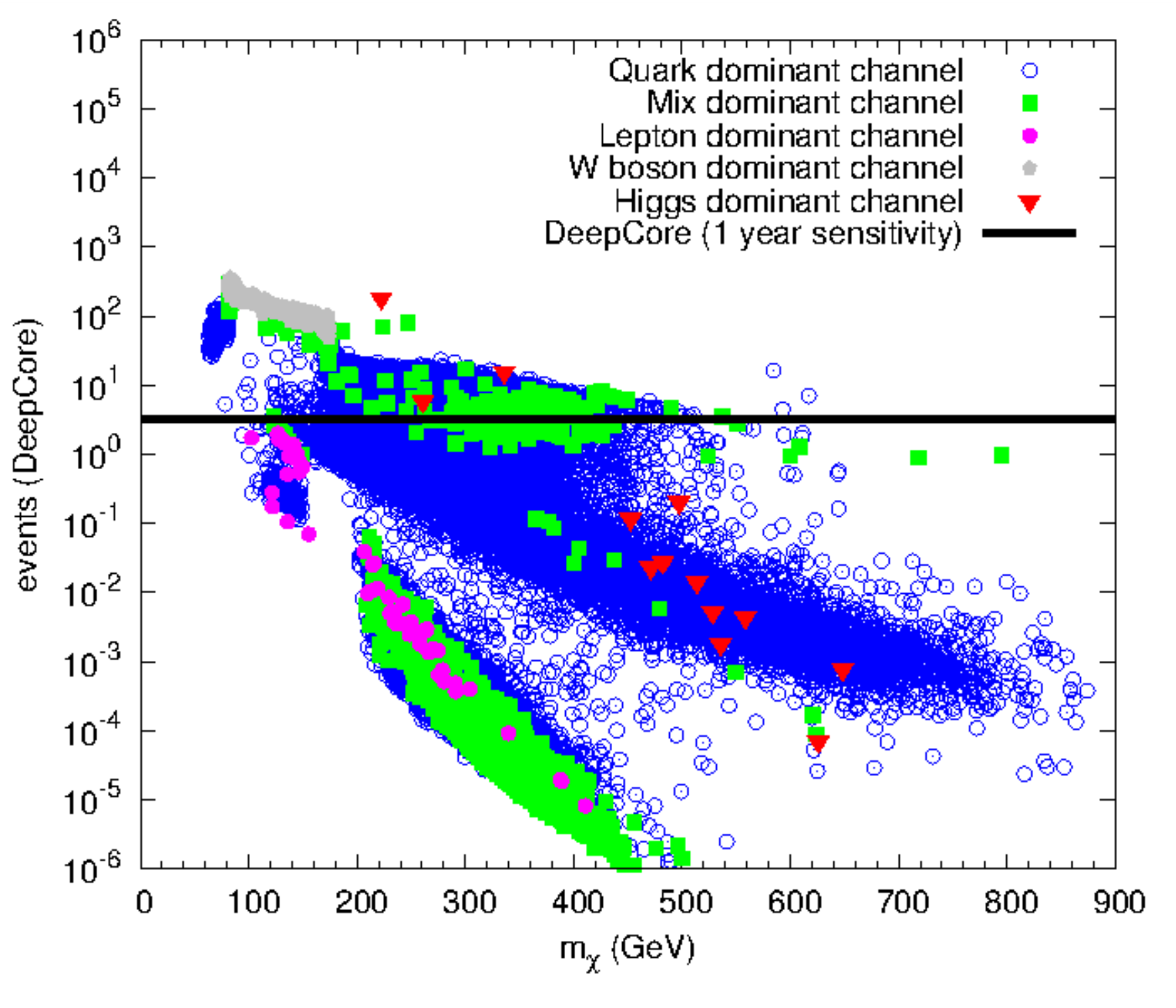}\\
\label{Group_m0_m12}%
\includegraphics[width=70mm,
height=70mm]{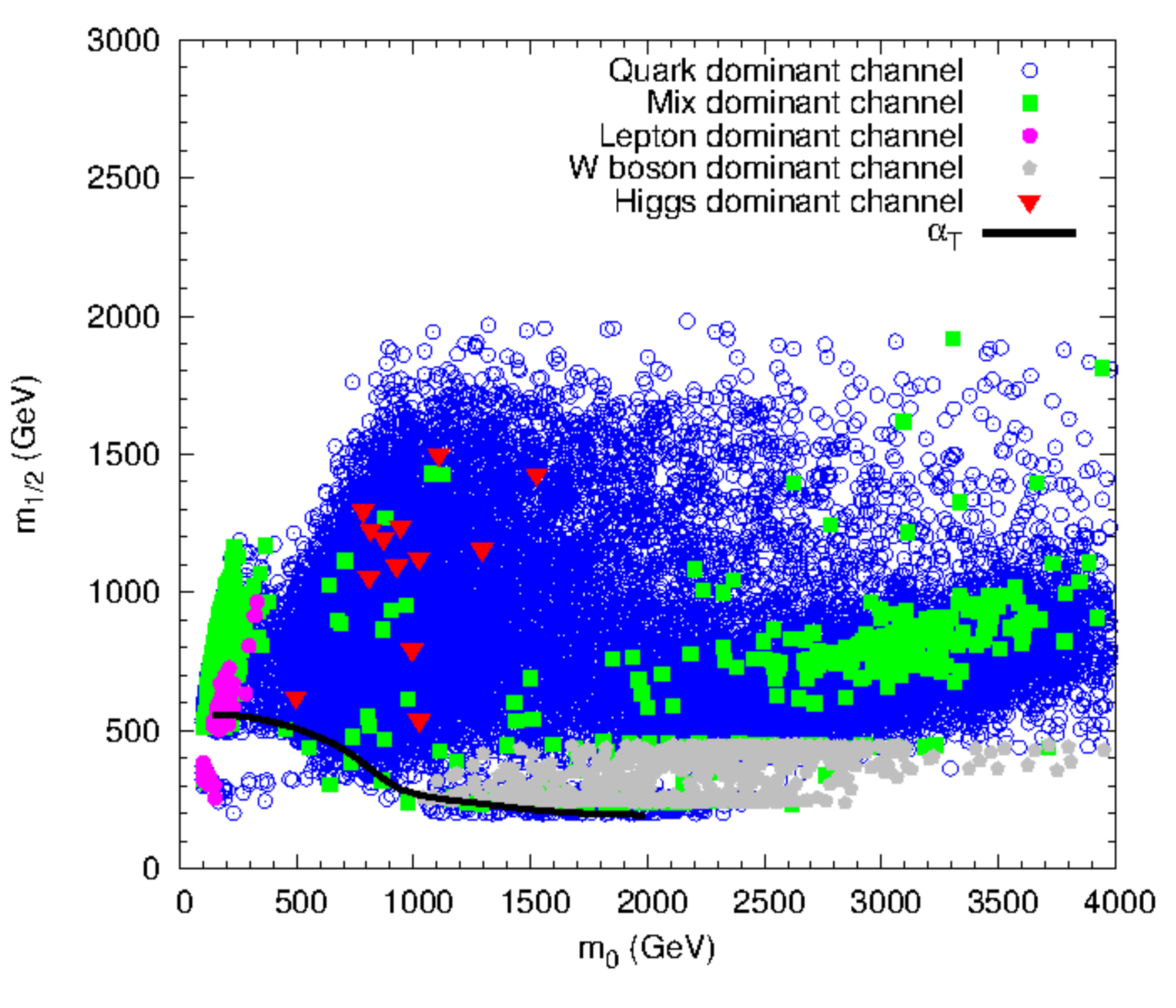}%
\label{PostFermi_mx_sigsip}%
\includegraphics[width=70mm,
height=70mm]{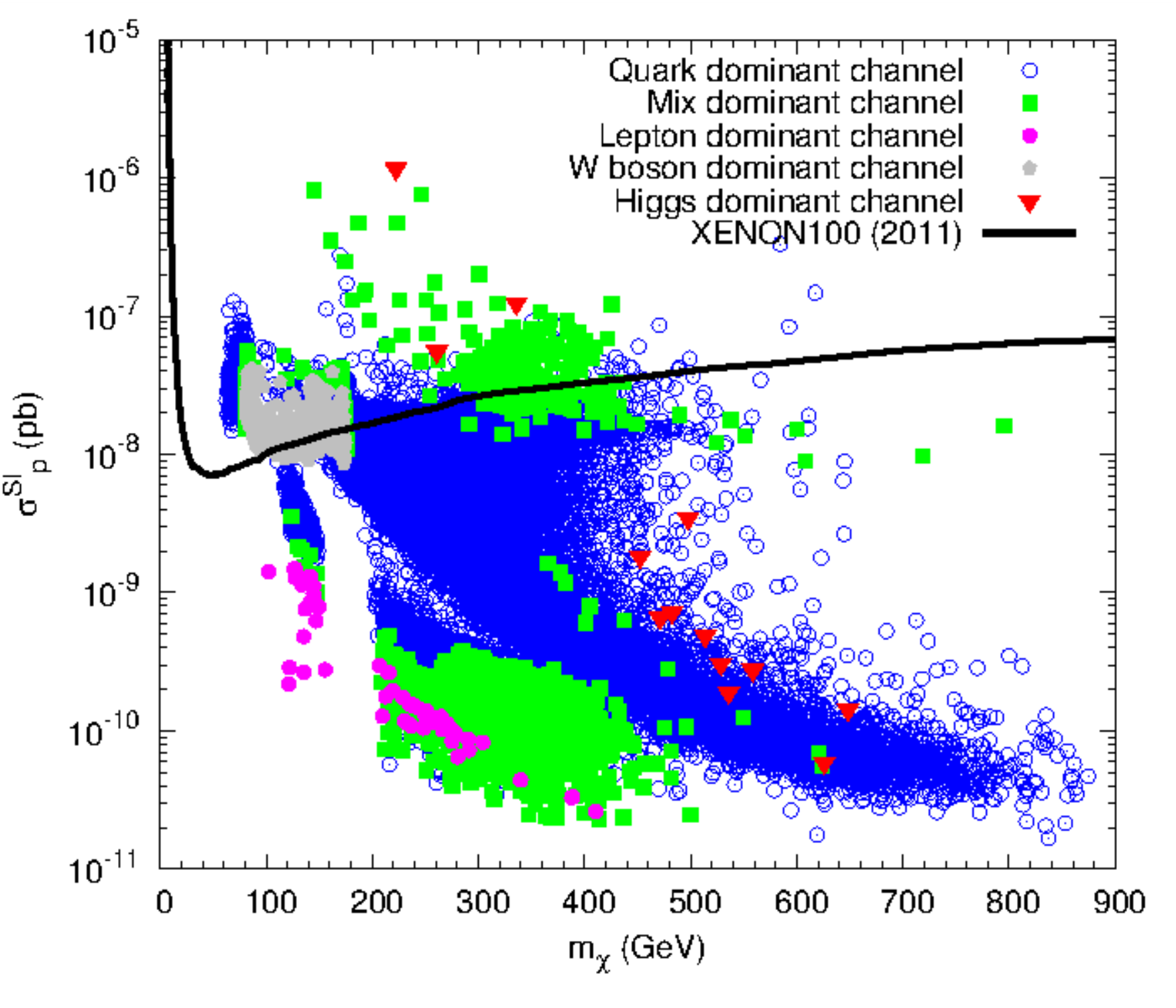}
\caption{The impact of some of the exclusion bounds and projected
  sensitivities considered in this work on models with a predominant
  neutralino annihilation channel. (a) Projected one-year
  95\%~C.L. sensitivity in contained events. (b) Projected
  sensitivity at DeepCore. (c) LHC $\alpha_T$
  bound. (d) XENON100 bound.} 
\end{figure}

Some very enticing possibilities arise in the case that IceCube/DeepCore lives up to its
expected discovery potential. In particular, we found that if the
projected sensitivities are achieved, and the experimental and
theoretical uncertanties are kept reasonably small, in one year
DeepCore could rule out the fraction of the FP region which is mostly
responsible for neutralino annihilation to $W^+W^-$ at the
95\%~C.L. To illustrate this point we show in Fig.~7 the impact of
some of the exclusion bounds and projected sensitivities considered in
this work on models with a single predominant neutralino annihilation
channel. Each point represents a case for which neutralinos
annihilate 90\% of the time to the selected final state, and for which
the bound on the relic abundance is respected
\textbf{($0.05<\Omega_{\chi}h^2<0.20$)}. The likelihood used to draw
the point distribution in space is subject to all of the constraints
discussed above. We show in
Fig.~7(a) the impact of the contained event
sensitivity (black horizontal line) on these models. It is more
constraining than the upward event sensitivity (not shown), but no
final state can be excluded in one year of observation. On the
other hand, Fig.~7(b) shows the same analysis
for DeepCore. One can see that the $WW$ channel could be excluded at the
95\%~C.L., and so could any CMSSM-like model for which the DM
annihilates into $W^+W^-$. The advantage of DeepCore comes from the
lower energy threshold. Since the $WW$ channel opens up around $m_W$,
which is very close to the threshold predicted for the IceCube's larger
volume, there is not enough phase space over which the muon flux can
be integrated in the contained events. Consequently, the number of hits
at the detector will not be sizeable. On the other hand, since the
muon energy threshold at DeepCore is $E_{\textrm{min}}\sim 35$~GeV, even neutralinos that
annihilate into the $WW$ channel can provide enough muon events to be
successfully detected. One can see in Fig.~7(c) that the
Higgs-resonance region and part of the FP region are mostly responsible for cases with predominant DM
annihilation to $W^+W^-$. We show in Fig.~7(d)
the corresponding analysis for the XENON100 exclusion bound on
$\sigma_p^{\textrm{SI}}$ (black line). A fraction of the cases that
annihilate to $W^+W^-$ are below the bound, so nothing definitive can
be said about the exclusion potential in any channel.

\section{Summary}\label{summary:sec}

In this paper we investigated the impact of the present bounds and
projected sensitivities from DM indirect detection searches on global
Bayesian inferences of the CMSSM. We applied the 95\%~C.L. upper bound
on $\langle\sigma v\rangle$ from the recent analysis of $\gamma$-ray
fluxes from dSphs by the FermiLAT Collaboration and calculated the
effect of the projected one-year 95\%~C.L. sensitivities for neutrino
fluxes from WIMP annihilation in the Sun at IceCube and DeepCore. We
also applied the constraints used in Ref.~\cite{Fowlie:2011mb}, with
respect to which we extended our scan deep into the FP region, up to
$m_0=4$~TeV and $m_{1/2}=2$~TeV. In particular, we considered the
recent lower bounds on sparticle masses from the CMS $\alpha_T$ search
with 1.1~fb$^{-1}$ integrated luminosity and the recent
90\%~C.L. upper bound on $\sigma_p^{\textrm{SI}}$ from XENON100. With
respect to~\cite{Fowlie:2011mb}, we improved the code for
determination of the efficiency maps by simulating the CMS detector
response with PGS4.

We found that, when the present constraints on the mass within the half-light radius are taken into account, the dSphs cross section bounds show significant constraining power on the FP/HB region. 
On the other hand,  we confirm the findings of our previous studies that, when the theoretical
uncertainties are fully taken into consideration, the constraining
power of the XENON100 bound is limited. However, a less
conservative (in fact, currently unrealistic) treatment of the
theoretical uncertainties involved, obtained often by including only the
experimental uncertainties (as given by the experimental
collaboration) shows significant impact on the same region of parameter space. Even more interestingly, the DM constraints considered here, both from DD and ID, become competitive in this case.
Finally, we showed that the projected one-year
sensitivity at DeepCore has the potential to further constrain the FP
region and to exclude the fraction of PS over which DM annihilates
predominantly to $W^+W^-$ final states. Thus, the next round of data
from IceCube/DeepCore has the potential to independently rule out any new physics
models whose coupling to SM particles is comparable in strength to the
CMSSM and whose products of annihilation are $W$ bosons.


\bigskip

\paragraph{Acknowledgments}
We would like to thank Kamila Kowalska for help during the revision of this paper. L.R. would like to thank Aldo Morselli, Eric Nuss and Alex Drlica-Wagner for useful discussions on the bounds from stellar velocity dispersions. 
  We are funded in part by the Welcome Programme
  of the Foundation for Polish Science. L.R. is also supported in part
  by the Polish National Science Centre Grant No. N202 167440, an STFC
  consortium grant of Lancaster, Manchester and Sheffield Universities
  and by the EC 6th Framework Program MRTN-CT-2006-035505.



\bibliographystyle{utphysmcite}	
\bibliography{myref}


\end{document}